\documentclass[aps, prl, twocolumn, superscriptaddress, amsmath,  tightenlines, longbibliography]{revtex4-1}

\usepackage{dcolumn}
\usepackage{graphicx}
\usepackage{mathrsfs}
\usepackage{subfigure}
\usepackage{booktabs}
\usepackage{amsmath}
\usepackage{physics}
\usepackage{dsfont}
\usepackage{amstext}
\usepackage{amssymb}
\usepackage{amsbsy}
\usepackage{bbm}
\usepackage{amsthm}
\usepackage{graphicx}
\usepackage{color}
\usepackage[colorlinks,citecolor=blue]{hyperref}

\usepackage{url}
\usepackage[colorlinks]{hyperref}
\hypersetup{%
	plainpages=true,
	breaklinks=true,       
	hypertexnames=false,  
	pageanchor=true,
	colorlinks=true,
	linkcolor={blue},
	citecolor={red},
	urlcolor={blue},
	anchorcolor={black}
}

 \makeatletter

\newcommand{\Rmnum}[1]{\expandafter\@slowromancap\romannumeral #1@}
\makeatother

\begin{document}

\title{Anderson Delocalization in Strongly-Coupled Disordered  Non-Hermitian Chains}
\author{Wei-Wu Jin}
\thanks{These authors contributed equally}
\affiliation{School of Physics and Optoelectronics, South China University of Technology,  Guangzhou 510640, China}
\author{Jin Liu}
\thanks{These authors contributed equally}
\affiliation{School of Physics and Optoelectronics, South China University of Technology,  Guangzhou 510640, China}
\author{Xin Wang}
\affiliation{Institute of Theoretical Physics, School of Physics, Xi'an Jiaotong University, Xi'an 710049,  China}
\author{Yu-Ran Zhang}
\affiliation{School of Physics and Optoelectronics, South China University of Technology,  Guangzhou 510640, China}
\author{Xueqin Huang}
\affiliation{School of Physics and Optoelectronics, South China University of Technology,  Guangzhou 510640, China}
\author{Xiaomin Wei}
\affiliation{School of Physics and Optoelectronics, South China University of Technology,  Guangzhou 510640, China}
\author{Wenbo Ju}
\email[E-mail: ]{wjuphy@scut.edu.cn}
\affiliation{School of Physics and Optoelectronics, South China University of Technology,  Guangzhou 510640, China}
\author{Zhongmin Yang}
\email[E-mail: ]{yangzm@scut.edu.cn}
\affiliation{School of Physics and Optoelectronics, South China University of Technology, Guangzhou 510640, China}
\affiliation{Research Institute of Future Technology, South China Normal University, Guangzhou 510006, China}
\affiliation{State Key Laboratory of Luminescent Materials and Devices and Institute of Optical Communication Materials, South China University of Technology, Guangzhou 510640, China}
\author{Tao Liu}
\email[E-mail: ]{liutao0716@scut.edu.cn}
\affiliation{School of Physics and Optoelectronics, South China University of Technology,  Guangzhou 510640, China}
\author{Franco Nori}
\affiliation{Center for Quantum Computing, RIKEN, Wakoshi, Saitama 351-0198, Japan}
\affiliation{Department of Physics, University of Michigan, Ann Arbor, Michigan 48109-1040, USA}

\date{{\small \today}}


\begin{abstract}
	Disorder and non‑Hermitian effects together can upend how waves localize. In a 1D disordered chain, the non‑Hermitian skin effect (NHSE) can induce Anderson delocalization, defying the usual rule that disorder in low dimensions always localizes states. While weak disorder leaves the NHSE intact, strong disorder restores Anderson localization. Here, we study a surprising twist: coupling a strongly disordered Hatano-Nelson chain to a disordered Hermitian chain with their disorder anti-symmetrically correlated. Strikingly, once the inter-chain coupling exceeds a threshold, the system undergoes Anderson delocalization irrespective of disorder strength, reinstating the NHSE with no Hermitian counterpart. This transition arises from the interplay of non-reciprocal hopping, inter-chain coupling, and engineered disorder correlations, and is captured by a real-space winding number. To confirm this, we build an electrical-circuit analog and directly observe the re-emergent NHSE via voltage measurements. Our work uncovers unexplored and experimentally accessible physics at the crossroads of non‑Hermiticity and disorder.
\end{abstract}

\maketitle

\textit{\color{blue}Introduction}.---Recent years have seen growing interest in exotic physics emerging from non-Hermitian systems \cite{PhysRevLett.118.040401, El-Ganainy2018, ShunyuYao2018, PhysRevLett.123.066404, PhysRevLett.122.076801, PhysRevLett.125.126402, PhysRevLett.124.086801, PhysRevLett.123.066405, PhysRevB.100.054105, Zhao2019, PhysRevX.9.041015, PhysRevLett.124.056802, PhysRevB.102.235151, PhysRevLett.127.196801, PhysRevB.105.205402, PhysRevLett.129.093001, Ren2022, PhysRevA.109.063329,PhysRevB.106.085427, arXiv:2408.12451,PhysRevX.13.021007,PhysRevB.111.205418, PhysRevAL061701,arXiv:2411.10398}. These include platforms such as optical setups  \cite{Regensburger2012, PhysRevLett.113.053604, Hodaei975, Peng2014, Peng328, Leefmans2022,  Parto2023, Leefmans2024NP, Zhang2018}, electrical circuits \cite{Choi2018, Helbig2020, wu2022non, Zou2021, Hu2023,PhysRevB.111.014304}, and open quantum systems   \cite{PhysRevLett.123.123601, PhysRevLett.124.147203}. A central discovery is the   non-Hermitian skin effect (NHSE) \cite{ShunyuYao2018, PhysRevLett.123.066404, PhysRevLett.122.076801,  PhysRevLett.125.126402},  where bulk modes become highly sensitive to boundaries and localize at the edges under open boundary conditions. This effect, rooted in point-gap topology\cite{PhysRevLett.125.126402,  PhysRevLett.124.086801}, drives many novel phenomena without Hermitian analogs, such as the breakdown of conventional Bloch band theory \cite{ShunyuYao2018} and disorder-free entanglement phase transitions \cite{PhysRevX.13.021007}.

Disorder plays a crucial role in non-Hermitian systems, impacting transport, entanglement  and topology \cite{PhysRevLett.77.570, PhysRevB.58.8384, PhysRevE.59.6433, PhysRevX.8.031079, PhysRevLett.126.166801, PhysRevB.100.054301, PhysRevB.101.165114, PhysRevLett.122.237601, PhysRevB.101.014202, PhysRevB.103.L140201, PhysRevLett.126.090402, PhysRevLett.127.213601, Weidemann2021, Lin2022, PhysRevB.107.144204}. It induces exotic phenomena such as nonunitary scaling in localization \cite{PhysRevLett.126.166801}, disorder-induced non-Bloch topological phase transitions \cite{Lin2022}, and coexistence of dynamical delocalization and spectral localization \cite{Weidemann2021}. A seminal study by Hatano and Nelson extended the one-dimensional (1D) Anderson model by introducing nonreciprocal hopping \cite{PhysRevLett.77.570}, revealing unexpected delocalization with the emergence of NHSE under weak disorder. This challenges the conventional expectation  that disordered systems in less than two dimensions are always localized \cite{PhysRevLett.42.673}. Their findings showed that non-Hermiticity can fundamentally alter Anderson localization in the 1D Hatano-Nelson (HN) chain. However, as disorder strength increases, the skin modes eventually localize again, highlighting a complex interplay between disorder and non-Hermiticity.  This reentrant  behavior raises a compelling question: can a strongly disordered non-Hermitian system exhibit an Anderson localization-delocalization transition that revives the NHSE? 

In Hermitian systems, coupling between disordered chains has been shown to induce delocalization \cite{PhysRevLett.81.862,PhysRevLett.96.076603,PhysRevB.90.155411,PhysRevLett.116.140401,PhysRevLett.123.036403,PhysRevA.109.033310,arXiv:2307.01638}, suggesting a potential route to explore similar transitions in the non-Hermitian field. Inspired by this,  we explore the Anderson localization-delocalization behavior of coupled non-Hermitian chains in a ladder geometry subjected to ultra-strong disorder. Our study aims  to uncover novel mechanisms driving Anderson localization-delocalization transitions  that revives the NHSE, offering deeper insights into the intricate interplay among non-Hermiticity, disorder, and inter-chain coupling in extended systems.

In this work, we uncover a novel topological Anderson localization-delocalization transition in coupled disordered chains, driven by the interplay of non-Hermiticity, inter-chain coupling, and correlated disorder. Our hybrid system consists of a non-Hermitian HN chain coupled to a Hermitian chain with anti-symmetrically correlated disorder. Remarkably, this configuration leads to an Anderson delocalization transition that revives the NHSE even under ultra-strong disorder, with the inter-chain coupling strength controlling the re-emergence of the NHSE. This transition is topologically characterized by the real-space winding number. Experimental realization using tunable nonreciprocal electrical circuits confirms these findings, highlighting how non-Hermiticity and correlated disorder together govern wave localization.

\textit{\color{blue}Model}.--- We start by considering a  HN chain with asymmetric hopping \cite{PhysRevLett.77.570}, which shows the NHSE. Adding random onsite disorder leads to Anderson delocalization at weak disorder \cite{PhysRevLett.77.570,PhysRevB.58.8384, PhysRevE.59.6433,PhysRevLett.126.166801,PhysRevX.8.031079}. As the disorder strength increases further, Anderson localization eventually takes over.  In this work, we demonstrate that Anderson delocalization, accompanied by re-emergent NHSE,  can be restored even in the ultra-strong disorder regime by coupling the disordered HN chain to a disordered Hermitian chain with correlated disorder, as illustrated in Fig.~\ref{Fig1}(a). The Hamiltonian of the hybrid system  is written as
\begin{align}\label{H1}
	\hat{\mathcal{H}}   = &~ \sum_{j} \left[ (\gamma +\lambda ) \hat{a}_{j}^\dagger  \hat{a}_{j+1} +  (\gamma -\lambda )\hat{a}_{j+1}^\dagger  \hat{a}_{j}  \right]  \nonumber \\
	& ~ + \sum_{j} \left(J  \hat{b}_{j+1}^\dagger  \hat{b}_{j}+ t  \hat{a}_{j}^\dagger  \hat{b}_{j} + \textrm{H.c.} \right)
	 \nonumber \\
	& ~+ \sum_{j} \left(\Delta_{j}^{(a)} \hat{a}_{j}^\dagger  \hat{a}_{j} + \Delta_{j}^{(b)} \hat{b}_{j}^\dagger  \hat{b}_{j}  \right)  ,
\end{align}
where $\hat{a}^\dagger_j$ and $\hat{b}^\dagger_j$ are   creation operators for the HN and Hermitian chains   at $j$th unit cell,   $\gamma \pm\lambda$ denote  the asymmetric hopping, $J$ is the symmetric hopping strength, $t$ is the inter-chain coupling strength,   and $\Delta_{j}^{(\mu)}$ ($\mu = {a,\, b}$)  is the random onsite  potential, applied to the HN ($\mu=a$) and   Hermitian   ($\mu=b$) chains, which is uniformly sampled in $[-W/2, W/2]$, with $W$ being the disorder strength.

\begin{figure}[!tb]
	\centering
	\includegraphics[width=8.5cm]{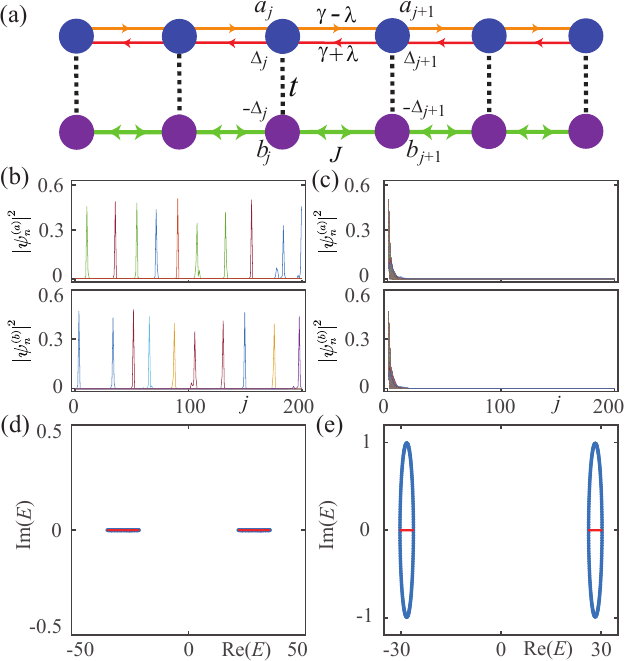}
	\caption{(a) Schematic showing a disordered HN chain (top) coupled to a disordered Hermitian chain  (bottom). $\gamma \pm \lambda$ denote the asymmetric hopping strengths, $J$ is the symmetric hopping strength,  and $t$ is the   inter-chain coupling strength. The anti-symmetric random onsite potential $\Delta_{j}$ and $-\Delta_{j}$ are applied to each chain.  (b,c) Probability density distributions $|\psi_{n}^{(a)}(j)|^2$ and $|\psi_{n}^{(b)}(j)|^2$ of the coupled chains subject to symmetric  random potential with  $\Delta_{j}^{(a)} = \Delta_{j}^{(b)} = \Delta_{j}$ (b), and anti-symmetric  random potential with  $\Delta_{j}^{(a)} = -\Delta_{j}^{(b)} = \Delta_{j}$ (c). All eigenstates exhibit Anderson localization, where only a few are plotted in (b) for clarity from one disorder realization. Plots in (d,e) show the corresponding complex eigenenergies under PBC (blue dots) and OBC (red dots). The parameters used are  $\gamma/J = \lambda/J =1$,  $W/J= 12$,  $t/J=28$, and $N=200$. 
	}\label{Fig1}
\end{figure}

Although uncorrelated strong random disorder leads to Anderson localization in both the HN chain and the coupled chains, we will show that the re-emergence of the NHSE can occur if an appropriately correlated disorder is applied to the coupled chains. We consider two types of correlated disorder schemes: symmetric disorder with $\Delta_{j}^{(a)} = \Delta_{j}^{(b)} = \Delta_{j}$, and anti-symmetric disorder with $\Delta_{j}^{(a)} = -\Delta_{j}^{(b)} = \Delta_{j}$.

Figure \ref{Fig1}(b-e) plots the probability density distributions $|\psi_{n}^{(a)}(j)|^2$ and $|\psi_{n}^{(b)}(j)|^2$ of right eigenstates under ultra-strong onsite  disorder, and the corresponding complex eigenenergies under periodic (PBC) and open (OBC) boundary conditions. In the presence of symmetrically correlated disorder, all states in the coupled HN-Hermitian chains remain localized [see Fig.~\ref{Fig1}(b)]. The absence of point gaps in Fig.~\ref{Fig1}(d) further indicates the breakdown of the NHSE. However, when anti-symmetrically correlated disorder is introduced, the complex eigenspectrum  under PBC (blue dots) forms two point gaps at large inter-chain coupling strength $t$ [see Fig.~\ref{Fig1}(e)], which encircle the eigenenergies  under OBC (red dots). As shown in Fig.~\ref{Fig1}(c), all states inside the point gaps are localized at the left boundary, signaling the re-emergence of the NHSE. Note that the eigenstates in the coupled Hermitian chains remain localized despite anti-symmetric disorder. This highlights that the Anderson delocalization accompanied by re-emergent NHSE in a disordered nonreciprocal chain, coupled to a Hermitian chain with anti-symmetric disorder, is a genuinely nontrivial phenomenon without a Hermitian counterpart.

The hidden physical mechanism can be intuitively understood by considering the strong inter-chain coupling case with $\abs{t} \gg \abs{J}, \abs{\gamma \pm \lambda}, \abs{\Delta_{j}}$ for both symmetric and anti-symmetric disorder configurations (see Sec.~\Rmnum{1} of the Supplementary Material (SM) \cite{SMBCS2024}). In this limit, rewriting $\hat{\mathcal{H}}$ in the new basis $\ket{\alpha_\pm,j } = \hat{\alpha}_{\pm, j}^\dagger \ket{0}$, with $\hat{\alpha}_{\pm,j} = (\pm \hat{a}_j + \hat{b}_j)/\sqrt{2} $, yields an effective nonreciprocal Creutz ladder subject to onsite random potentials $ \Delta_{j} \pm  t$ and  $\pm \sqrt{t^2+\Delta_{j}^2 }$ for symmetric and anti-symmetric disorder configurations, respectively. For symmetric disorder, the effective Creutz ladder undergoes Anderson localization under strong disorder $\Delta_{j}$. However, for anti-symmetric disorder, the effective disorder strength $\tilde{W}$ becomes $\tilde{W} < W^2/\abs{8t}$ \cite{SMBCS2024}. In the strong inter-chain coupling limit $\abs{t} \gg W$, $\tilde{W}$ becomes much smaller than the nonreciprocal hopping strength $\lambda$ of the Creutz ladder. While an arbitrarily small amount of disorder induces Anderson localization   in 1D Hermitian systems \cite{PhysRevLett.42.673} (see also Sec.~\Rmnum{4} of SM \cite{SMBCS2024}), the interplay between nonreciprocal hopping and disorder can instead  lead to an Anderson transition \cite{PhysRevLett.77.570,PhysRevX.8.031079}. Consequently, increasing inter-chain coupling $t$ drives the Anderson delocalization with the re-emergent NHSE, even under arbitrarily strong anti-symmetric disorder.

To further characterize the Anderson localization-delocalization transition induced by the anti-symmetrically correlated disorder, we calculate the inverse participation ratio (IPR) of each normalized right eigenstate $\psi_{n} = (\psi_{n}^{(a)}, \psi_{n}^{(b)})^T$,  defined as
\begin{align}
	\textrm{IPR}_{n} =  \sum_{j=1}^N \left(| \psi_{n}^{(a)}(j)|^4  + | \psi_{n}^{(b)}(j)|^4   \right).
\end{align}\label{eq:IPRR}
Here, $\textrm{IPR}_{n} \simeq 1/(2N)$ for an extended eigenstate $\psi_{n}$ and vanishes as $N \to \infty$, while for a state localized over $M \ll N$ sites, $\textrm{IPR}_{n} \simeq 1/M$ and remains finite in the thermodynamic limit.

\begin{figure}[!tb]
	\centering
	\includegraphics[width=8.7cm]{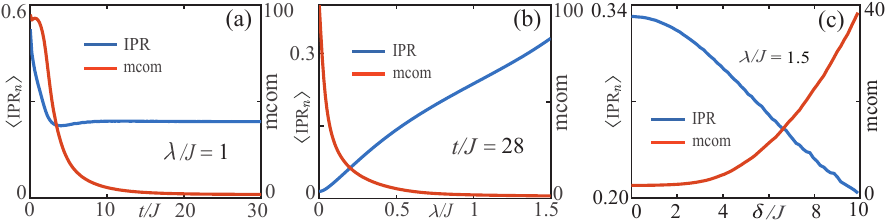}
	\caption{(a-c) IPR and mcom averaged over all eigenstates, under anti-symmetric disorder, (a) as a function  of  $t$  with   $\lambda/J=1$,   (b) as a function  of  $\lambda$ with  $t/J=28$, and   (c) as a function  of   $\delta$ with   $\lambda/J=1.5$, where a random perturbation $\delta_{j} \in [-\delta/2,~\delta/2]$ breaks the exact anti-symmetry. The results  are averaged over $2000$ disorder realizations with $\gamma/J =1$, $ W/J=12$ and $N=200$.} \label{Fig2}
\end{figure}

However, the IPR cannot distinguish skin-mode localization from Anderson localization. To resolve skin modes, we compute the mean center of mass (mcom) of the squared amplitudes of all right eigenvectors $\psi_{n}$, averaged over disorder realizations \cite{PhysRevResearch.5.033058}, defined as
\begin{align}
	\mathrm{mcom}  = \frac{\sum_{j=1}^N j \left< \mathcal{A}(j) \right>}{\sum_{j=1}^N \left< \mathcal{A}(j) \right>} , 
\end{align}\label{eq:mcom}
with the mean amplitude squared of all right eigenstates 
\begin{align}
	\left< \mathcal{A}(j) \right> &= \left< \frac{1}{2N} \sum_{n=1}^{2N} \left(| \psi_{n}^{(a)}(j)|^2  + | \psi_{n}^{(b)}(j)|^2   \right) \right>,
\end{align}
where $\left< \cdot \right>$  denotes averaging over disorder realizations.  When mcom is close to be $1$ or $N$, it indicates that  the eigenstates are localized at the boundaries, with the emergence of NHSE. As explained in Sec.~\Rmnum{2} of the SM \cite{SMBCS2024}, the IPR and mcom averaged over all eigenmodes effectively resolve delocalized and extended states and distinguish skin-mode localization from Anderson localization. In addition, to avoid finite-size effects (see details in Sec.~\Rmnum{3} of SM \cite{SMBCS2024}), we perform our calculations on a large lattice.

Figures \ref{Fig2}(a) and (b) show the averaged IPR and mcom under anti-symmetric disorder as functions of inter-chain coupling $t$ and asymmetric hopping $\lambda$, respectively. Strong disorder ($W/J=12$) induces Anderson localization at small $t$ [Fig.~\ref{Fig2}(a)], while increasing $t$ triggers Anderson delocalization and revives the NHSE with boundary-localized eigenstates. Furthermore, with fixed disorder ($W/J=12$) and hopping strength ($t/J=28$), Anderson delocalization occurs only when $\lambda$ exceeds a critical value, highlighting the necessity of strong nonreciprocal hopping and inter-chain coupling for the re-emergence of the NHSE under strong anti-symmetric disorder. Note that the IPR based on biorthogonal eigenvectors shows the same localization behavior as that using right eigenvectors alone, as shown in Sec.~\Rmnum{4} of SM \cite{SMBCS2024}. In addition, the re-emergent NHSE  can be further revealed through quenched evolution dynamics (see details in Sec.~\Rmnum{5} of SM \cite{SMBCS2024}).

We now investigate the robustness of this phenomenon under deviations from exact anti-symmetric disorder. To this end, we consider a modified disorder configuration with $\Delta_{j}^{(a)} = \Delta_{j}$ and $\Delta_{j}^{(b)} = -\Delta_{j} + \delta_{j}$, where $\delta_{j} \in [-\delta/2, \delta/2]$ is a random variable characterizing symmetry-breaking perturbations (see details in Sec.~\Rmnum{6} of SM \cite{SMBCS2024}). Figure \ref{Fig2}(c) plots the averaged IPR and mcom as a function  of $\delta$, showing that the NHSE persists even with deviations up to $25\%$ ($\delta=3$) from perfectly  anti-symmetric disorder configuration. This demonstrates the robustness of re-emergent NHSE against imperfect anti-symmetric disorder,  extending its applicability beyond fine-tuned scenarios.

\textit{\color{blue}Phase diagram}.---In order to determine the phase diagrams of  the Anderson localization to delocalization transitions with the emergence of skin modes, due to the triple interplay of anti-symmetric disorder, nonreciprocal hopping and inter-chain coupling, we calculate the winding number and mcom for different parameters. The winding number in real space is defined as \cite{PhysRevB.103.L140201}
\begin{align}
	w(E_b) = \frac{1}{N'} \text{Tr}' \left( \hat{Q}^{\dagger} [\hat{Q},\hat{X}] \right),
\end{align}
where $\hat{Q}$ is a positive-definite Hermitian matrix, which is obtained by the polar decomposition $(\hat{\mathcal{H}}-E_b) = \hat{Q}\hat{P}$, with unitary matrix $\hat{P}$. $\hat{Q}$ and $\hat{P}$ are related to the singular value decomposition $(\hat{\mathcal{H}}-E_b) = \hat{M}\hat{S}\hat{N}^\dagger$, with $\hat{Q} = \hat{M}\hat{N}^\dagger$ and $\hat{P} = \hat{N}\hat{S}\hat{N}^\dagger$. $\hat{X}$ is the coordinate operator, with $X_{jj',ss'}= j\delta_{j,j'}\delta_{s,s'} (s=a,b)$, and $\text{Tr}'$ denotes the trace over the middle interval with length $N'$, where the whole chain is cut off from both ends. This definition of the winding number avoids the effects from the system's boundary.

\begin{figure}[!tb]
	\centering
	\includegraphics[width=8.7cm]{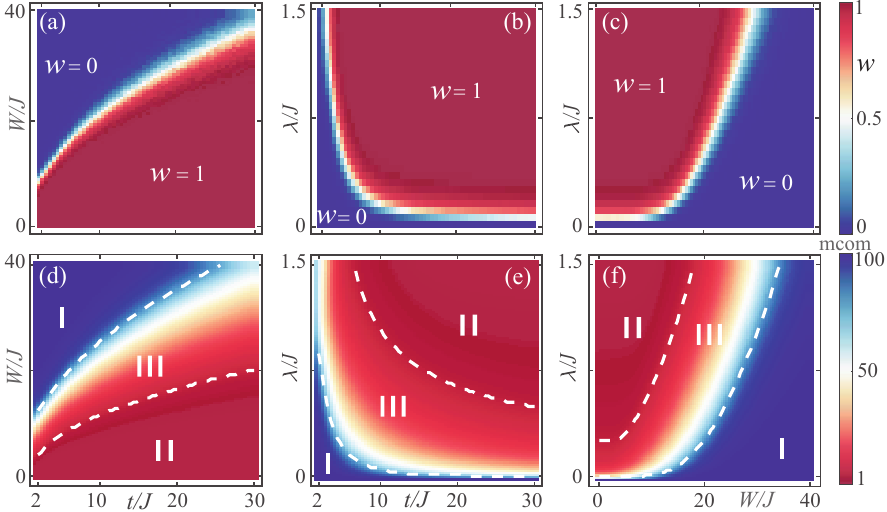}
	\caption{(a-c) Winding number $w$ and (d-f) mcom, under anti-symmetric disorder, in the $(W, t)$ plane with $\lambda/J =1$ (a,d), the $(\lambda, t)$ plane  with $W/J=12$ (b,e), and the $(\lambda, W)$ plane  with  $t/J=14$ (c,f). Regions \Rmnum{1}, \Rmnum{2}, and \Rmnum{3} correspond to Anderson localization, skin-mode localization,  and a mixture of the two, respectively. The results  are averaged over $1200$ disorder realizations with $\gamma/J =1$ and $N=200$.  } \label{Fig3}
\end{figure}

The phase diagrams in the presence of anti-symmetric disorder are shown in Fig.~\ref{Fig3}(a-c), where the phase boundary between the absence ($w=0$) and presence ($w=1$) of skin modes is clearly visible. We find that, regardless of the disorder strength $W$, skin modes reappear once the inter-chain coupling $t$ becomes sufficiently large [see Fig.~\ref{Fig3}(a)]. Moreover, the phase boundary strongly depends on the asymmetric hopping strength $\lambda$ [see Fig.~\ref{Fig3}(b,c)]. The dependence of the phase diagrams on $\gamma$ is discussed in Sec.~\Rmnum{7} of SM \cite{SMBCS2024}.
	
The winding number cannot identify the coexistence regions of skin-mode localization and Anderson localization (see details in Sec.~\Rmnum{8} of SM \cite{SMBCS2024}), whereas this distinction can be made using the mcom, as shown in Fig.~\ref{Fig3}(d–f). Regions \Rmnum{1}, \Rmnum{2}, and \Rmnum{3} correspond to Anderson localization, skin-mode localization,  and a mixture of the two, respectively. Moreover, the phase boundary (white dotted line across region \Rmnum{1}) between the absence and presence of skin modes, determined by mcom, agrees with that obtained from real-space winding number.

\begin{figure}[!tb]
	\centering
	\includegraphics[width=8.7cm]{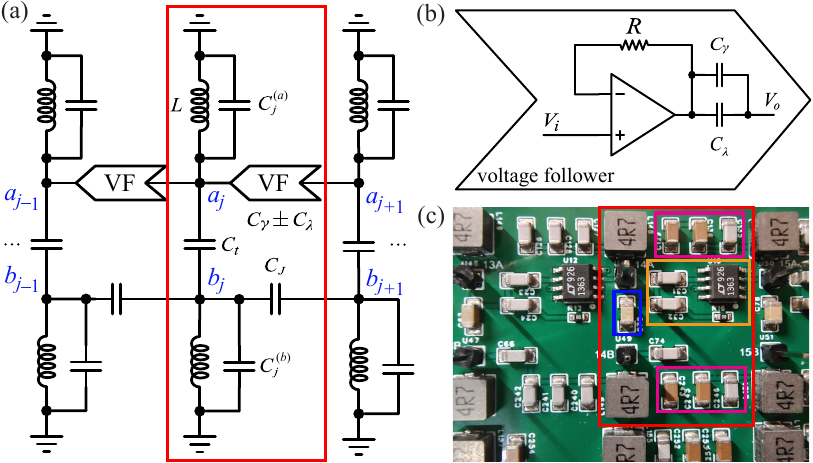}
	\caption{(a)  Circuit implementation of the coupled HN-Hermitian lattice subject to the correlated disorder, corresponding to the  model in Fig.~\ref{Fig1}(a). The red solid line outlines a unit cell in the circuit lattice. For the HN chain $\{a_j\}$, the nonreciprocal intra-chain hopping, represented by capacitors $C_{\gamma} \pm C_{\lambda}$, is realized  via a voltage follower (VF) with its circuit diagram shown in (b), where the resistor $R=1~\text{k}\Omega$ is used to ensure its stability. The capacitor $C_J$ denotes the intra-chain hopping in the Hermitian chain $\{b_j\}$, $C_t$ represents the inter-chain hopping. $C_j^{(a)}$ and $C_j^{(b)}$ are disordered capacitances to simulate the  correlated disorder. The inductor $L$ is used to adjust the resonance  frequency of the circuit.  (c) Photograph of the fabricated electric circuit in the enlarged view of a unit cell outlined by the red solid line. The modules outlined by blue, yellow and pink solid lines denote the capacitor $C_t$, VF and capacitor $C_J$, respectively. } \label{Fig4}
\end{figure}

\textit{\color{blue}Experimental implementation}.---We implement electrical circuits to realize Anderson delocalization with re‐emergent NHSE in coupled HN-Hermitian chains induced by correlated disorder. Figure~\ref{Fig4}(a) shows the designed circuit network, where the unit cell containing sublattices $a_j$ and $b_j$ is outlined by the red solid line. In the HN chain ${a_j}$, the nonreciprocal hopping between nodes $j$ and $j+1$, represented by capacitors $C_{\gamma} \pm C_{\lambda}$, is implemented using a voltage follower \cite{wu2022non} [see its circuit in Fig.~\ref{Fig4}(b)]. In the Hermitian chain ${b_j}$, the hopping is realized by capacitor $C_J$. $C_t$ denotes the inter-chain coupling. The correlated disorder is introduced through disordered capacitances $C_j^{(a)}$ and $C_j^{(b)}$. Figure \ref{Fig4}(c) shows the fabricated circuit board.
 
\begin{figure*}[!tb]
	\centering
	\includegraphics[width=18cm]{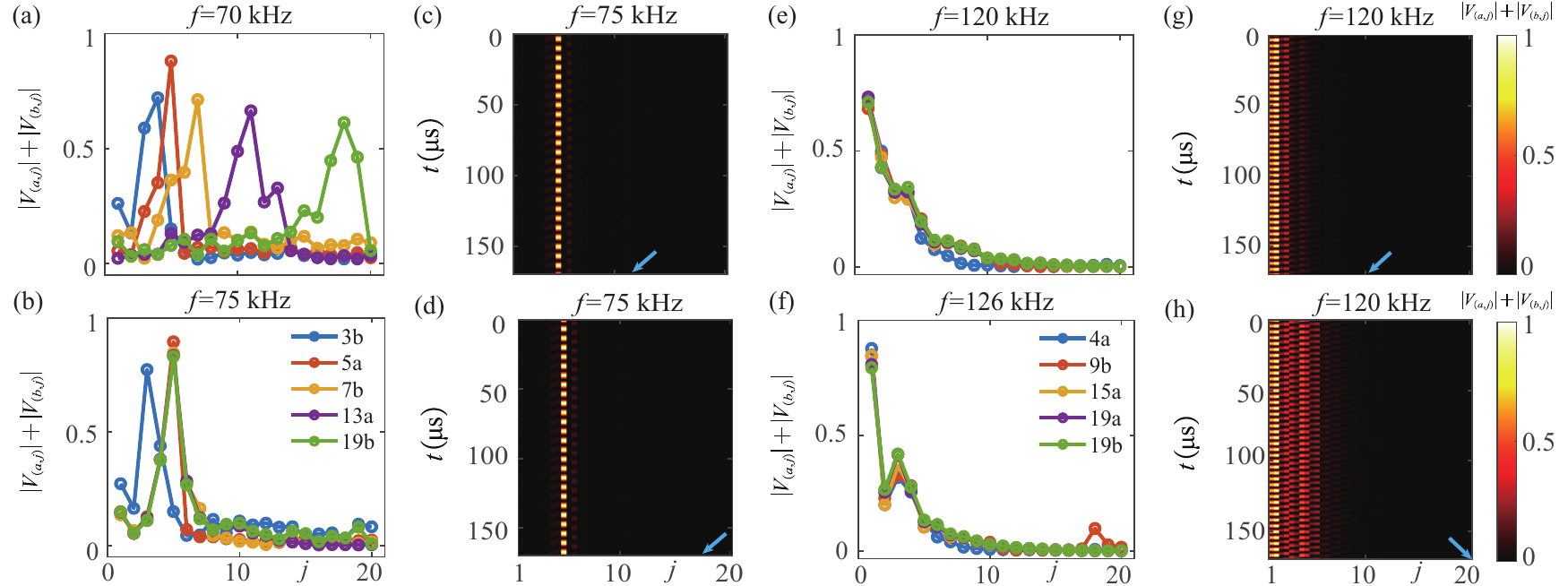}
	\caption{(a,b) Measured site-resolved voltage distributions $\abs{V_{a,j}}+\abs{V_{b,j}}$ ($j$ is unit-cell index)  at resonance frequencies $f=70 ~\textrm{kHz}$ (a) and $f=75 ~\textrm{kHz}$ (b) for weak inter-chain  hopping $C_t$ with $C_{\gamma}=C_{\lambda}=C_J=C_t=100 ~\text{nF}$,  $L=4.7~\mu\text{H}$, and $C_{j}^{\left( a \right)} = -C_{j}^{\left( b \right)} \in [-6 C_J,~6 C_J]$. The legend ``$j\alpha$" ($\alpha=a,b$) in (b) indicates the excitation at the $j$th site of the chain $\alpha$. (c,d) Measured temporal voltage responses excited at the $11$th (c) and $17$th (d) unit cells, indicated by the blue arrows,   for the weak inter-chain  hopping with $f=75 ~\textrm{kHz}$.   (e,f) Measured voltage distributions  and (g,h) temporal voltage responses   for the strong inter-chain  hopping $C_t$ with $C_{\gamma}=C_{\lambda}=C_J=47~\text{nF}$, $C_t=1~\mu\text{F}$,   $L=4.7~\mu\text{H}$, and $C_{j}^{\left( a \right)} = -C_{j}^{\left( b \right)} \in [-6 C_J,~6 C_J]$. } \label{Fig5}
\end{figure*}

The model in Eq.~(\ref{H1}) is represented by the   Laplacian $J(\omega)$ of the circuit, which is defined as the response of the voltage vector  $\mathbf{V}$ to the input current vector $\mathbf{I}$  by $\mathbf{I}(\omega) = J(\omega)\mathbf{V}(\omega)$. As shown in Fig.~\ref{Fig4}(a), the circuit Laplacian $J(\omega)$ reads (see details   in Sec.~\Rmnum{9} of SM \cite{SMBCS2024})
\begin{align}\label{eq410}
	J(\omega) = i\omega\mathcal{H} _{\textrm{c}}-\left( 2i\omega C_J+i\omega C_t+i\omega C_g+\frac{1}{i\omega L} \right) \mathds{1},
\end{align}
where $\mathds{1}$ is the $2N \times 2N$ identity matrix, and the matrix $\mathcal{H} _{\textrm{c}}$ is shown in Sec.~\Rmnum{9} of SM \cite{SMBCS2024}. 

The circuit Laplacian $J(\omega)$ and $\hat{\mathcal{H}}$ in Eq.~(\ref{H1}) share the same eigenstates  if we set $C_J = J$,   $C_t =t$, $C_J+C_\lambda = \gamma+\lambda$, $C_J-C_\lambda = \gamma-\lambda$, and $	  C_{j}^{\left( \alpha \right)} = - \Delta_{j}^{\left( \alpha \right)}$ ($\alpha=a,b$). The eigenstates of $J(\omega)$, corresponding to the circuit frequency $\omega = 2 \pi f$, can be obtained by measuring the voltage response at the circuit nodes.

Figure \ref{Fig5}(a,b)  shows the measured  site-resolved  voltage distributions  $\abs{V_{a,j}}+\abs{V_{b,j}}$, subject to  the anti-symmetric disorder. We resonantly excite the circuit at different nodes with resonance frequencies   $f=70 ~\textrm{kHz}$  and $f=75 ~\textrm{kHz}$, respectively, and then measure the voltages at all the nodes. For   weak inter-chain hopping strength $C_t$, we observe the Anderson-localized voltages distributions [see Fig.~\ref{Fig5}(a,b)].  This localization is further confirmed by temporal voltage measurements excited at the 11th and 17th unit cells at $f=75~\textrm{kHz}$ [see Fig.\ref{Fig5}(c,d)], where the voltages remain localized over time, consistent with the steady-state distributions in Figs.~\ref{Fig5}(a,b).

We increase the capacitance $C_t$ to enhance the inter-chain coupling and excite the circuit at frequencies $f=120~\textrm{kHz}$ and $f=126~\textrm{kHz}$, respectively. The measured site-resolved voltage distributions $\abs{V_{a,j}}+\abs{V_{b,j}}$ [see Fig.~\ref{Fig5}(e,f)], excited at different nodes under anti-symmetric disorder,  reveal re-emergent skin effects, consistent with theoretical predictions. In contrast, voltages remain localized under symmetrically correlated disorder (see details in Sec.~\Rmnum{10} of SM \cite{SMBCS2024}). Temporal voltage measurements [see Fig.~\ref{Fig5}(g,h)], excited at the 11th and 20th unit cells at  frequency $f=120~\textrm{kHz}$,  show voltage propagation along the left boundary, confirming the Anderson delocalization with the re‐emergent NHSE. These results experimentally demonstrate the anti-symmetrical disorder-induced re-emergent NHSE in a nonreciprocal chain coupled to a Hermitian chain.  Our circuit simulations further confirm the robustness of re-emergent NHSE against imperfections in the anti-symmetric disorder, which are inevitable in practical systems, thereby extending its applicability beyond fine-tuned conditions.

\textit{\color{blue}Conclusion}.---In summary, we have shown that coupled HN and Hermitian chains with anti-symmetrically correlated disorder exhibit  a topological Anderson localization–delocalization transition and a robust NHSE, phenomena absent in Hermitian systems. While an isolated disordered HN chain remains localized, coupling to a disordered Hermitian chain with anti-symmetric disorder induces an Anderson delocalization. Notably, strong inter-chain coupling universally leads to Anderson delocalization, accompanied by the re‐emergent NHSE, regardless of the disorder strength. The topological character of this transition is captured by a real-space winding number. Our electrical–circuit realization reproduces these effects in site-resolved voltage measurements, in excellent agreement with theoretical predictions.

Furthermore, our theoretical and experimental results demonstrate that Anderson delocalization with re-emergent NHSE remains robust even in the presence of imperfect anti-symmetric disorder. This indicates that  correlated disorder can drive Anderson delocalization and stabilize the NHSE in strongly disordered systems, extending beyond finely tuned conditions. This proof-of-concept highlights that, when combined with inter-chain coupling, correlated disorder serves as a tunable control parameter to induce transitions between localized and delocalized phases and to re-establish the NHSE. This tunability opens avenues for applications in quantum switches and devices, where wave transport  properties can be dynamically modulated via inter-chain coupling. Looking ahead, it will be compelling to extend this framework to higher-dimensional disordered non-Hermitian systems, where symmetry may play a crucial role in shaping Anderson transitions. Additionally, incorporating interactions could reveal new insights into the interplay of correlated disorder, non-Hermiticity, and many-body effects in governing localization and non-Hermitian phenomena.

\begin{acknowledgments}
 T.L. gratefully acknowledges helpful discussions with Zhongbo Yan. T.L. conceived and initiated the study and acknowledges the support from  the National Natural
Science Foundation of China (Grant No. 12274142), the Fundamental Research Funds for the Central Universities (Grant No.~2023ZYGXZR020), Introduced Innovative Team Project of Guangdong Pearl River Talents Program (Grant No. 2021ZT09Z109),  and the Startup Grant of South China University of Technology (Grant No.~20210012). Y.R.Z. thanks the support from the National Natural Science Foundation of China (Grant No.~12475017) and Natural Science Foundation of Guangdong Province (Grant No.~2024A1515010398). W.J. thanks the support from the National Natural Science Foundation of China (No.~U21A2093) and Introduced Innovative Team Project of Guangdong Pearl River Talents Program (Grant No.~2021ZT09Z109). 
F.N. is supported in part by: the Japan Science and Technology Agency (JST) [via the CREST Quantum Frontiers program Grant No. JPMJCR24I2, the Quantum Leap Flagship Program (Q-LEAP), and the Moonshot R$\&$D Grant Number JPMJMS2061].
\end{acknowledgments}

W.W.J. and J.L. contributed equally to this work.

\textit{Data availability}.---The data that support the findings of
this Letter are openly available \cite{zenodo.16411878}.


%

\clearpage \widetext
\begin{center}
	\section*{Supplemental Material for ``Anderson Delocalization in Strongly-Coupled Disordered  Non-Hermitian Chains"}
\end{center}
\setcounter{equation}{0} \setcounter{figure}{0}
\setcounter{table}{0} \setcounter{page}{1} \setcounter{secnumdepth}{3} \makeatletter
\renewcommand{\theequation}{S\arabic{equation}}
\renewcommand{\thefigure}{S\arabic{figure}}
\renewcommand{\bibnumfmt}[1]{[S#1]}
\renewcommand{\citenumfont}[1]{S#1}

\makeatletter
\def\@hangfrom@section#1#2#3{\@hangfrom{#1#2#3}}
\makeatother

\maketitle

\section{Physical mechanism} 

To gain intuitive insight into the physical mechanism behind Anderson delocalization accompanied by the re-emergence of the non-Hermitian skin effect (NHSE) under anti-symmetrically correlated disorder, we consider the regime of strong inter-chain coupling. Specifically, we assume $\abs{t} \gg \abs{J}, \abs{\gamma \pm \lambda}, \abs{\Delta_{j}}$, for both symmetric and anti-symmetric disorder configurations. In this limit, inter-chain hopping dominates the dynamics, allowing us to treat the intra-chain hopping 
terms as perturbations.  The  Hamiltonian $\hat{\mathcal{H}}$ is thus decomposed as $\hat{\mathcal{H}}_\pm = \hat{\mathcal{H}}_{0,\pm} + \hat{\mathcal{V}}$, with the  unperturbed part reading 
\begin{align}\label{unpertur}
	\hat{\mathcal{H}}_{0,\pm} = \sum_{j} \left[  t  \left(  a_{j}^\dagger  \hat{b}_{j} + \textrm{H.c.} \right)  + \Delta_{j} \left(a_{j}^\dagger  a_{j} \pm  \hat{b}_{j}^\dagger  \hat{b}_{j}  \right)\right],
\end{align}
and the perturbed part as
\begin{align}\label{pertur}
	\hat{\mathcal{V}}   = &~ \sum_{j} \left[ (\gamma +\lambda ) \hat{a}_{j}^\dagger  \hat{a}_{j+1} +  (\gamma -\lambda )\hat{a}_{j+1}^\dagger  \hat{a}_{j}  \right]   + \sum_{j} J \left( \hat{b}_{j+1}^\dagger  \hat{b}_{j} + \textrm{H.c.} \right),
\end{align}
where $\hat{\mathcal{H}}_+$ denotes the coupled chains under the symmetric disorder configuration, and $\hat{\mathcal{H}}_-$  under the anti-symmetric disorder configuration. 

In the strong inter-chain coupling limit, with $\abs{t} \gg \abs{J}, \abs{\gamma \pm \lambda}, \abs{\Delta_{j}}$, $\mathcal{H}_{0,\pm}$ can be rewritten as 
\begin{align}\label{unpertur2}
	\hat{\mathcal{H}}_{0,\pm} = \sum_{j,m=\pm} \xi_{\pm,j}^{(m)} \hat{\alpha}_{m,j}^\dagger \hat{\alpha}_{m,j}, 
\end{align}
where $\hat{\alpha}_{\pm,j} $ and  $\xi_{\pm,j}^{\pm}$ are 
\begin{align}\label{unpertur21}
	\hat{\alpha}_{\pm,j} = (\pm \hat{a}_j + \hat{b}_j)/\sqrt{2} , 
\end{align}
\begin{align}\label{unpertur22}
	\xi_{+,j}^{\pm} = \Delta_{j}\pm t, ~~~\textrm{and}~~~ \xi_{-,j}^{\pm} = \pm \sqrt{t^2+\Delta_{j}^2 }.
\end{align}

In the new basis $\ket{\alpha_\pm,j } = \hat{\alpha}_{\pm, j}^\dagger \ket{0}$, we  write $\hat{\mathcal{H}}_\pm $ as $\hat{\tilde{\mathcal{H}}}_\pm = \hat{\mathcal{H}}_{0,\pm} + \hat{\mathcal{H}}_{\textrm{ladder}}$, with

	\begin{align}\label{new}
		\hat{\mathcal{H}}_{\textrm{ladder}} = & ~\frac{J+\gamma+\lambda}{2}\sum_{j} \left(\hat{\alpha}_{+, j}^\dagger \hat{\alpha}_{+,j+1} + \alpha_{-, j}^\dagger \hat{\alpha}_{-,j+1} \right) +	\frac{J+\gamma-\lambda}{2}\sum_{j} \left(\hat{\alpha}_{+, j+1}^\dagger \alpha_{+,j} + \hat{\alpha}_{-, j+1}^\dagger \hat{\alpha}_{-,j} \right) \nonumber \\ 
		&+	\frac{J-\gamma-\lambda}{2}\sum_{j} \left(\hat{\alpha}_{-, j}^\dagger \hat{\hat{\alpha}}_{+,j+1} + \hat{\alpha}_{+, j}^\dagger \hat{\alpha}_{-,j+1} \right) +	\frac{J-\gamma+\lambda}{2}\sum_{j} \left(\alpha_{+, j+1}^\dagger \hat{\alpha}_{-,j} + \hat{\alpha}_{-, j+1}^\dagger \hat{\alpha}_{+,j} \right).
	\end{align}

In Eq.~(\ref{new}), we have neglected  disorder in the hopping terms, which is quite small for $\abs{\Delta_{j}} \gg \abs{J}, \abs{\gamma \pm \lambda}$.

In this new basis, the Hamiltonian $\hat{\tilde{\mathcal{H}}}\pm$ takes the form of a nonreciprocal Creutz ladder, denoted $\hat{\mathcal{H}}_{\textrm{ladder}}$ \cite{PhysRevLett.83.2636SM}, subject to a disordered onsite potential $\hat{\mathcal{H}}{0,\pm}$. The disorder configuration depends on the symmetry of correlations: $\hat{\mathcal{H}}_{0,+}$ for symmetric and $\hat{\mathcal{H}}_{0,-}$ for anti-symmetric disorder. For the parameter regime considered in the main text, $\hat{\mathcal{H}}_{\textrm{ladder}}$ exhibits the non-Hermitian skin effect (NHSE), with all eigenmodes accumulating at the left boundary. In the symmetric disorder case, the effective onsite potential on each chain of the Creutz ladder is given by $\Delta_j \pm t$ ($\Delta_j \in [-W/2,~W/2]$ denotes random disorder). This  disorder  leads to the Anderson localization when the disorder strength $W$ becomes sufficiently large.

However, in the case of anti-symmetric disorder, the onsite random potential in each leg of the ladder takes the form $\pm \sqrt{t^2 + \Delta_j^2}$ ($\Delta_j \in [-W/2,~W/2]$ denotes random disorder). Under this configuration, the effective disorder strength $\tilde{W}$ in the Creutz ladder is bounded as $\tilde{W} < W^2 / |8t|$. In the strong inter-chain coupling limit, where $|t| \gg |\Delta_j|$ (i.e., $|t| \gg W$), this results in a disorder strength $\tilde{W}$ that is significantly weaker than the nonreciprocal hopping amplitude $\lambda$. While even infinitesimal disorder induces Anderson localization in one-dimensional Hermitian systems \cite{PhysRevLett.42.673SM}, it has been shown that the interplay between nonreciprocal hopping and disorder can instead lead to an Anderson transition \cite{PhysRevLett.77.570SM,PhysRevX.8.031079SM}. Consequently, as the inter-chain hopping $t$ increases, the NHSE can re-emerge, even in the presence of arbitrarily strong disorder, provided the disorder remains anti-symmetric.

\section{Eigenenergy-resolved $\textrm{IPR}$ and $\textrm{mcom}$}

In the main text, we utilize the IPR and mcom, averaged over all eigenstates and disorder realizations, to distinguish between skin-mode localization and Anderson localization. By averaging over the full spectrum, we capture the collective localization behavior of the system. Although the eigenstate ensemble may contain both localized and extended states, the averaged IPR reflects the system’s overall tendency toward localization or delocalization. In combination with the IPR, the averaged mcom serves as an effective diagnostic tool for identifying skin-mode localization, characterized by $\textrm{mcom} \sim 1$ or $\textrm{mcom} \sim N$, and for distinguishing it from Anderson localization, which typically yields $\textrm{mcom} \sim N/2$. Without this distinction, the two types of localization—skin-mode and Anderson—would be mixed.

\begin{figure*}[!t]
	\centering
	\includegraphics[width=17cm]{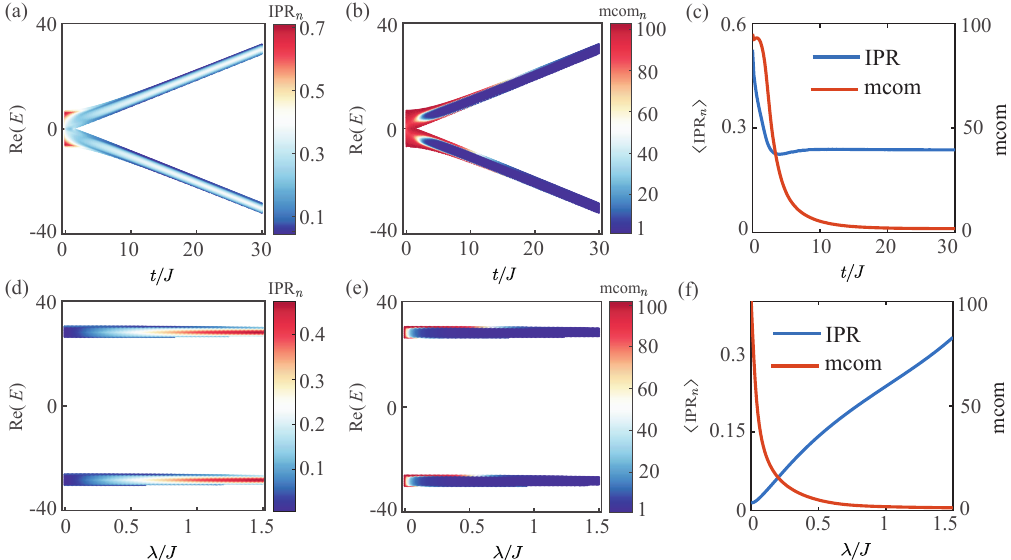}
	\caption{ Eigenenergy-resolved $\textrm{IPR}_n$ and $\textrm{mcom}_n$ of the coupled chain, subject to anti-symmetrically correlated disorder with $\Delta_{j}^{(a)} = -\Delta_{j}^{(b)} = \Delta_{j}$, plotted  as a function of coupling strength $t$ for $W/J = 12$ and $\lambda/J = 1$ (a,b), and  as a function of asymmetric hopping strength $\lambda$ for $W/J = 12$ and $t/J = 28$ (d,e). The corresponding IPR averaged over all the eigenstates   are shown in (c,f).  The remaining parameters are fixed at $\gamma/J = 1$ and $N = 200$. All results are averaged over $2000$ disorder realizations. }\label{FigS7}
\end{figure*}

In this section, we present eigenenergy-resolved results for $\textrm{IPR}_n$ and $\textrm{mcom}_n$, averaged over disorder realizations. These results demonstrate that IPR and mcom, averaged over all eigenstates and disorder realizations,  can indeed be effectively employed to detect localization and to distinguish between skin-mode and Anderson localization in the system.

Each skin mode can be identified by evaluating the eigenenergy-resolved mean center of mass $\textrm{mcom}_n$, which quantifies the spatial localization of the squared amplitude of each eigenstate, averaged over disorder realizations. The eigenenergy-resolved mcom   is defined as
\begin{align}
	\mathrm{mcom}_n  = \frac{\sum_{j=1}^N j \left< \mathcal{A}_n(j) \right>}{\sum_{j=1}^N \left< \mathcal{A}_n(j) \right>} , 
\end{align}\label{eq:mcom2}
with  $\left< \mathcal{A}_n(j) \right>$ being 
\begin{align}
	\left< \mathcal{A}_n(j) \right> &= \left<   \left(| \psi_{n}^{(a)}(j)|^2  + | \psi_{n}^{(b)}(j)|^2   \right) \right>,
\end{align}
where $\left< \cdot \right>$ indicates disorder averaging. When $\textrm{mcom}_n$ is close to $1$ or $N$, it indicates that the eigenstate is localized at the boundaries, signaling the emergence of the NHSE. In contrast, Anderson localization typically yields $\textrm{mcom} \sim N/2$, reflecting bulk localization.

Figure \ref{FigS7} shows the eigenenergy-resolved $\textrm{IPR}_n$  of the coupled chain, subject to anti-symmetrically correlated disorder with $\Delta_{j}^{(a)} = -\Delta_{j}^{(b)} = \Delta_{j}$, as functions of the inter-chain hopping strength $t$ in (a) and the asymmetric intra-chain hopping strength $\lambda$ in (d). The localized and extended states are clearly resolved by the  $\textrm{IPR}_n$: a finite value $\textrm{IPR}_n \sim 1/M$ (with $M \ll N$) indicates localization, while $\textrm{IPR}_n \sim 1/(2N)$ corresponds to extended states. These localization features are also reflected in the IPR averaged over all eigenstates and disorder realizations, as shown in Fig.~\ref{FigS7}(c,f). However, it is important to note that the IPR alone cannot distinguish between skin-mode localization and Anderson localization.

\begin{figure}[!t]
	\centering
	\includegraphics[width=18cm]{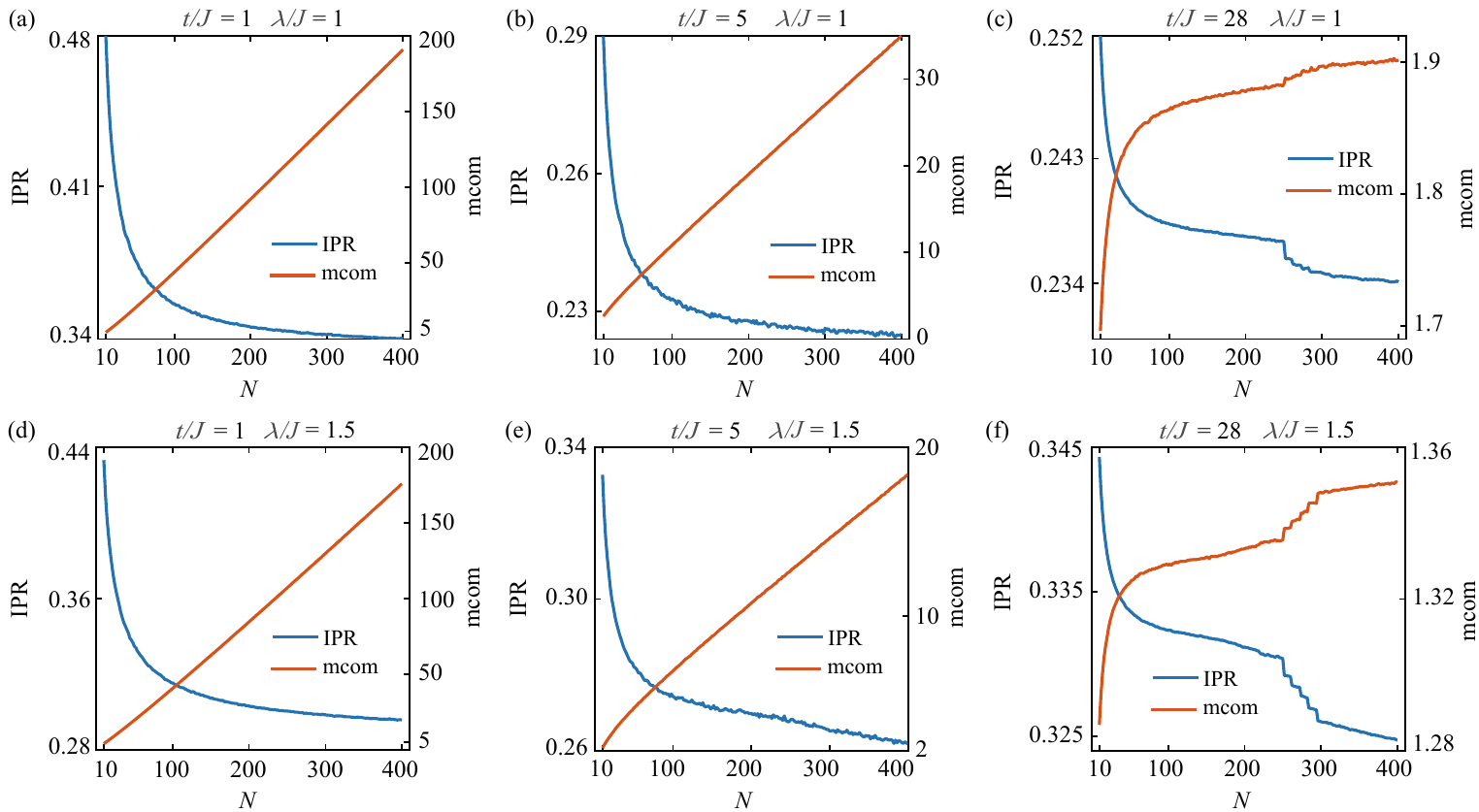}
	\caption{Impact of lattice size on Anderson localization and delocalization.  IPR and mcom    averaged over all the eigenstates, subject to anti-symmetric disorder, as a function of lattice size $N$   (a) for $t/J =1$ and $\lambda/J =1$, (b) for $t/J =5$ and $\lambda/J =1$, (c) for $t/J =28$ and $\lambda/J =1$, (d) for $t/J =1$ and $\lambda/J =1.5$, (e) for $t/J =5$ and $\lambda/J =1.5$, and (f) for $t/J =28$ and $\lambda/J =1.5$. The results  are averaged over $1200$ disorder realizations with $\gamma/J =1$ and $W/J=12$. } \label{FigS6}
\end{figure}

In order to distinguish skin-mode localization from Anderson localization, we calculate the eigenenergy-resolved $\textrm{mcom}_n$ as functions of the inter-chain hopping strength $t$ [see Fig.~\ref{FigS7}(b)] and the asymmetric intra-chain hopping strength $\lambda$ [see Fig.~\ref{FigS7}(e)]. As either $t$ or $\lambda$ increases, the system undergoes a transition: initially, all eigenstates are Anderson localized, characterized by $\textrm{mcom}_n \sim N/2$. Then, the system enters a regime where skin-mode localization and Anderson localization coexist. Eventually, all states become skin modes with $\textrm{mcom}_n \sim 1$. These distinct regimes are also captured by the $\textrm{mcom}$ averaged over all eigenstates [see Fig.~\ref{FigS7}(c,f)]: $\textrm{mcom} \sim N/2$ indicates Anderson localization, $\textrm{mcom} \sim 1$ signals skin-mode localization, and intermediate values  reflect the coexistence of both localization types.

\section{Finite-size Effects }

In the main text, the Anderson localization-delocalization transition is investigated using a fixed lattice size. In this section, we examine the impact of varying the lattice size $N$ on the localization behavior and demonstrate that the conclusions presented in the main text remain unchanged as the lattice size is further increased.

Figure \ref{FigS6} shows the IPR and mcom, averaged over all eigenstates and subject to anti-symmetric disorder, as functions of the lattice size $N$ for different inter-chain coupling strengths $t$  and asymmetric intra-chain hopping amplitudes $\lambda$. For Anderson-localized phases with $\textrm{mcom} \sim N/2$ [see Fig.~\ref{FigS6}(a,d)] and for skin-mode localization with $\textrm{mcom} \sim 1$ [see Fig.~\ref{FigS6}(c,f)], the phase identification remains robust and largely independent of system size. In both cases, the mcom serves as a clear and reliable phase indicator. However, in the coexistence regime of Anderson and skin-mode localization [see Fig.~\ref{FigS6}(b,e)], the mcom becomes less effective for small system sizes ($N < 100$), as its value remains low and does not distinguish the mixed phase well from pure skin-mode localization. To resolve this phase more clearly, the mcom should be evaluated in larger systems. In contrast, the IPR provides a consistent indicator of localization properties even for a small-size system size.

\section{Biorthogonal inverse participation ratio}

In the main text, we use the inverse participation ratio (IPR) based on right eigenstates to characterize localization properties. As an alternative, one can also consider the biorthogonal inverse participation ratio (BIPR) \cite{PhysRevB.105.075128SM}, defined as
\begin{align}
	\textrm{BIPR}_{n} =  \frac{\sum_{j=1}^N \left(| \psi_{n,L}^{(a)}(j)|^2  | \psi_{n,R}^{(a)}(j)|^2  + | \psi_{n,L}^{(b)}(j)|^2 |\psi_{n,R}^{(b)}(j)|^2   \right) }{\left(\sum_{j=1}^N | \psi_{n,L}^{(a)}(j)|  | \psi_{n,R}^{(a)}(j)|  + | \psi_{n,L}^{(b)}(j)| |\psi_{n,R}^{(b)}(j)|   \right)^2},
\end{align}
where $\psi_{n,L}^{(a)}$ and $\psi_{n,R}^{(a)}$ ($\psi_{n,L}^{(a)}$ and $\psi_{n,R}^{(a)}$) are left and right eigenstates in chain $a$ ($b$) .  Here, $\textrm{BIPR}_{n} \simeq 1/(2N)$ for an extended eigenstate $\psi_{n}$ and vanishes as $N \to \infty$, while for a state localized over $M \ll N$ sites, $\textrm{BIPR}_{n} \simeq 1/M$ and remains finite in the thermodynamic limit.

\begin{figure*}[!t]
	\centering
	\includegraphics[width=18cm]{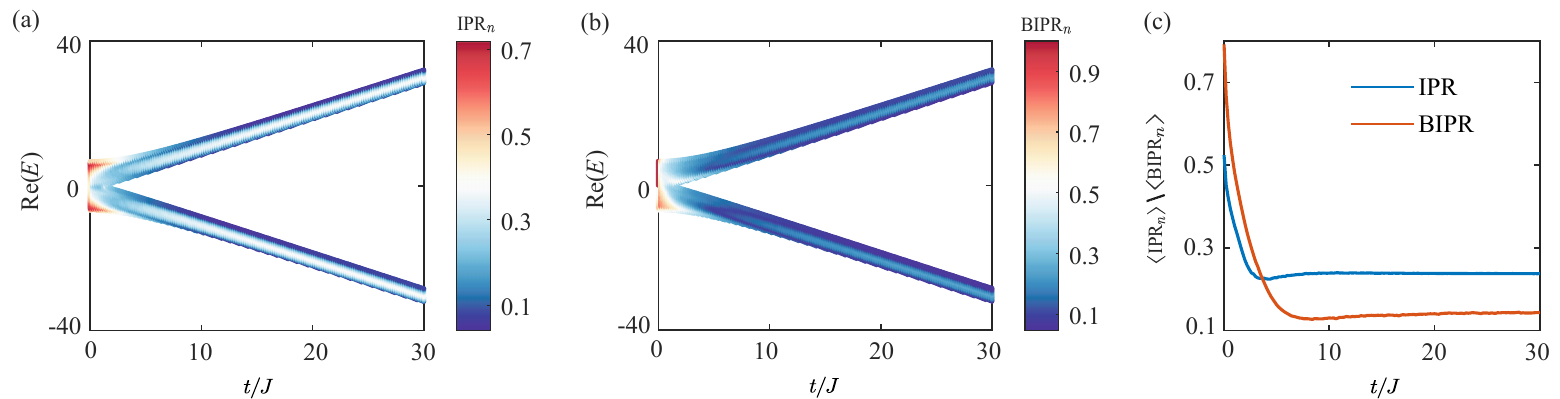}
	\caption{ Eigenenergy-resolved $\textrm{IPR}_n$ (a) and $\textrm{BIPR}_n$ (b) of the coupled chain, subject to anti-symmetrically correlated disorder with $\Delta_{j}^{(a)} = -\Delta_{j}^{(b)} = \Delta_{j}$, plotted  as a function of coupling strength $t$. The corresponding IPR and BIPR averaged over all the eigenstates   are shown in (c).  The remaining parameters are fixed at $W/J = 12$, $\gamma/J = 1$ and $N = 200$. All results are averaged over $1200$ disorder realizations. }\label{FigS9}
\end{figure*}

Figure \ref{FigS9} presents the eigenenergy-resolved $\textrm{IPR}_n$ (a) and $\textrm{BIPR}_n$ (b) for the coupled chain under anti-symmetrically correlated disorder, plotted as functions of the coupling strength $t$. The corresponding IPR and BIPR averaged over all eigenstates are shown in panel (c). While the values of the IPR and BIPR differ slightly, both consistently indicate that the system remains in a localized phase, as evidenced by their finite values. This confirms that either measure can effectively characterize localization. Therefore, using the IPR based on the right eigenstates alone is sufficient to distinguish localized states from extended ones.

\section{Quenched dynamics}

\begin{figure}[!tb]
	\centering
	\includegraphics[width=16.1cm]{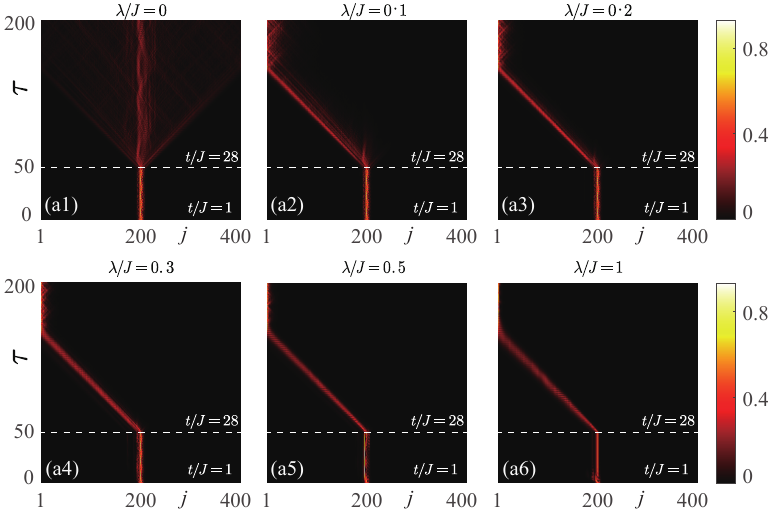}
	\caption{Dynamical localization and delocalization via quench. Quenched dynamics of density distributions for different $\lambda$  subjected to anti-symmetric disorder, where the initial state is set as the Gaussian wavepacket as $\psi_0(j) = \exp[-(j-j_0)^2/2\sigma^2]/\mathcal{N}$ centered at the site $j_0$ in the HN chain. At time $\tau=50$, the  hopping strength changes from $t/J=1$ to $t/J=28$. The   parameters used are $\gamma/J =1$, $W/J = 12$, and $N=400$ with (a1) $\lambda/J=0$, (a2) $\lambda/J=0.1$, (a3) $\lambda/J=0.2$, (a4) $\lambda/J=0.3$, (a5) $\lambda/J=0.5$, and (a6) $\lambda/J=1$.}\label{FigS1}
\end{figure}

The Anderson delocalization accompanied by re-emergent skin modes induced by anti-symmetric disorder can be further examined through the study of quenched evolution dynamics. The initial state is chosen as a Gaussian wavepacket as $\psi_0(j) = \exp[-(j-j_0)^2/2\sigma^2]/\mathcal{N}$ centered at the site $j_0$, where $\mathcal{N}$ is the normalization constant, and $\sigma$ denotes the wavepacket width. The wavefunction at time $\tau$ is obtained by numerically calculating $\ket{\psi(j,\tau)} = \exp(-i \mathcal{H}\tau) \ket{\psi_0(j)}$. Figure \ref{FigS1} plots quenched dynamics of density distributions for different $\lambda$  subjected to anti-symmetric disorder, where  the  hopping strength suddenly changes from $t/J=1$ to $t/J=28$ at time $\tau=50$. For the Hermitian case with $\lambda/J = 0$ [see Fig.~\ref{FigS1}(a1)], the initial localized mode remains mostly localized after the quench. The slight  spreading of the density distribution is attributed to finite-size effects. As the asymmetric hopping parameter $\lambda$  increases, the wavepacket initially localized at the center of the ladder becomes delocalized, and propagates towards the left boundary after the quench, and it is finally  localized at the boundary, due to the interplay of the NHSE, inter-chain coupling  and anti-symmetric disorder.

\begin{figure}[!t]
	\centering
	\includegraphics[width=18cm]{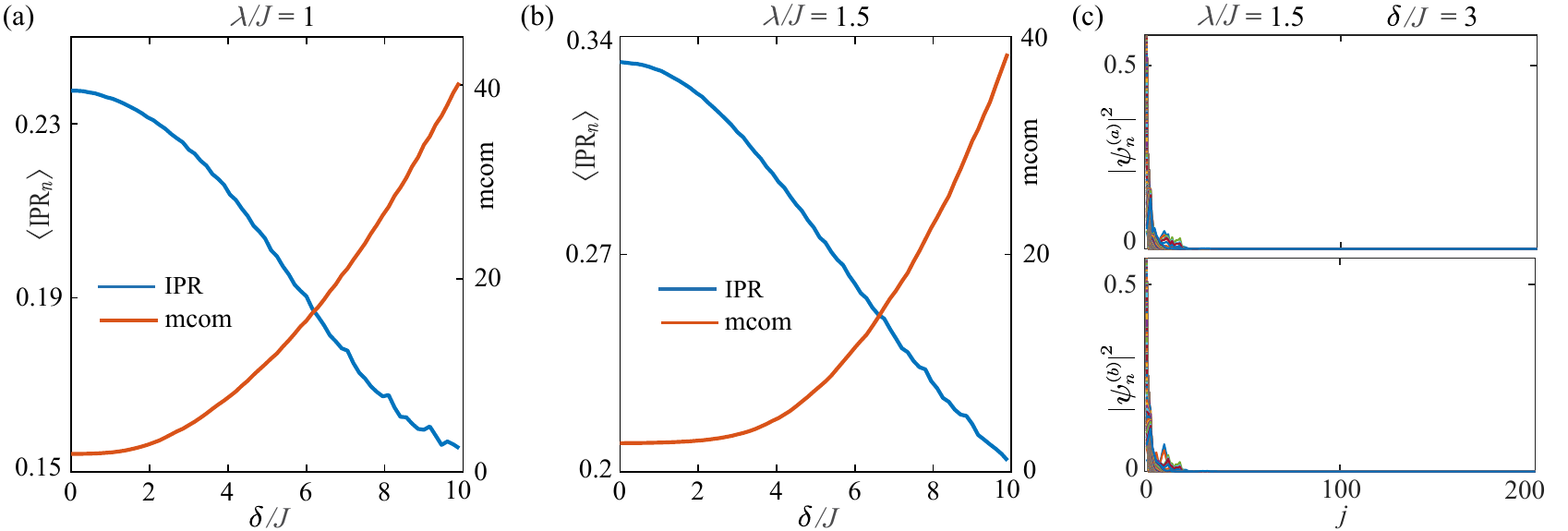}
	\caption{Impact of imperfect anti-symmetric correlated disorder on Anderson delocalization. (a,b) IPR and mcom averaged over all the eigenstates, subject to anti-symmetric disorder $\Delta_{j}$ and a random perturbation $\delta_{j}$ that breaks the exact anti-symmetry, as a function of the perturbation strength $\delta$ for $\lambda/J =1$ (a) and  $\lambda/J =1.5$ (b). The results  are averaged over $2000$ disorder realizations with $\gamma/J =1$, $W/J=12$, $t/J = 28$, and $N=200$.  (c) Probability density distributions $|\psi_{n}^{(a)}(j)|^2$ and $|\psi_{n}^{(b)}(j)|^2$ of all the eigenstates  subject to anti-symmetric disorder $\Delta_{j}$ and a random perturbation $\delta_{j}$ for $\lambda/J =1.5$  and  $\delta/J =3$ in one disorder realization with $N=200$. } \label{FigS5}
\end{figure}

\section{Robustness of Anderson Delocalization Against Imperfect Anti-Symmetric Disorder}

In the main text, we demonstrated that anti-symmetric disorder, defined by $\Delta_{j}^{(a)} = -\Delta_{j}^{(b)} = \Delta_{j}$, can give rise to Anderson delocalization in the coupled Hermitian and non-Hermitian chains. This Anderson delocalized phase allows the NHSE to reappear, even in the presence of ultra-strong disorder. In this section, we investigate how robust this phenomenon is when the anti-symmetric condition is no longer exact. To this end, we consider a modified disorder configuration where $\Delta_{j}^{(a)} = \Delta_{j}$ and $\Delta_{j}^{(b)} = -\Delta_{j} + \delta_{j}$, with $\delta_{j}$ being a random variable that quantifies deviations from perfectly anti-symmetry. Then, the Hamiltonian of the hybrid system  reads
\begin{align}\label{H1}
	\hat{\mathcal{H}}_\textrm{dis}   = &~ \sum_{j} \left[ (\gamma +\lambda ) \hat{a}_{j}^\dagger  \hat{a}_{j+1} +  (\gamma -\lambda )\hat{a}_{j+1}^\dagger  \hat{a}_{j}  \right]    + \sum_{j} \left(J  \hat{b}_{j+1}^\dagger  \hat{b}_{j}+ t  \hat{a}_{j}^\dagger  \hat{b}_{j} + \textrm{H.c.} \right)
	\nonumber \\
	& ~+ \sum_{j} \Delta_{j} \left(  \hat{a}_{j}^\dagger  \hat{a}_{j} -   \hat{b}_{j}^\dagger  \hat{b}_{j}  \right)  + \sum_{j} \delta_{j}   \hat{b}_{j}^\dagger  \hat{b}_{j},
\end{align}
where $\Delta_{j}$ represents the anti-symmetric disorder applied to both the non-Hermitian  and Hermitian chains. It is uniformly sampled from the interval $[-W/2, ~W/2]$, where $W$ denotes the disorder strength. The term $\delta_{j}$ introduces a random perturbation that breaks the exact anti-symmetry, and is independently drawn from a uniform distribution in $[-\delta/2, ~\delta/2]$, with $\delta$ characterizing the degree of asymmetry in the disorder.

Figures \ref{FigS5}(a) and (b) show the mcom and    IPR, averaged over all eigenstates, in the presence of anti-symmetric disorder $\Delta_{j}$ and a random perturbation $\delta_{j}$ that explicitly breaks the anti-symmetry. These quantities are plotted as functions of the perturbation strength $\delta$ for $\lambda/J = 1$ in panel (a) and $\lambda/J = 1.5$ in panel (b), for two different values of the asymmetric hopping strength. In both cases, the finite value of the IPR indicates that the eigenstates of the coupled system remain localized, either due to Anderson localization or localization induced by the NHSE.

When the disorder deviates moderately from the anti-symmetric configuration, i.e., when the perturbation $\delta_j$ is of moderate strength, the eigenstates stay localized near the left boundary due to the re‐emergent NHSE. The small value of mcom in Fig.~\ref{FigS5}(a) confirms the persistence of this re‐emergent NHSE for deviations as large as approximately $16.7\%$ at $\lambda/J = 1$, despite the presence of strong disorder $\Delta_{j}$. Furthermore, for $\lambda/J = 1.5$, the NHSE persists even with a deviation of about $25\%$ from the anti-symmetric disorder configuration, despite the presence of strong disorder, as shown in Fig.~\ref{FigS5}(b). For moderate values of $\delta$, the eigenstates manifest as skin modes localized at the left boundary. As $\delta$ increases, a coexistence emerges between skin-mode localization and Anderson localization. With a further increase in $\delta$, the system undergoes a transition into a regime dominated by strong perturbation-induced Anderson localization, where the eigenstates become localized in the bulk.

Figure \ref{FigS5}(c) shows the probability density distributions $|\psi_{n}^{(a)}(j)|^2$ and $|\psi_{n}^{(b)}(j)|^2$ for all eigenstates, obtained from a single disorder realization with anti-symmetric disorder $\Delta_{j}$ and a random perturbation $\delta_{j}$, with $\lambda/J = 1.5$ and $\delta/J = 3$. These distributions reveal the   Anderson delocalization accompanied by the  NHSE even with a deviation of   $25\%$ from the anti-symmetric disorder configuration.   

These results demonstrate  that the Anderson delocalization, accompanied by the re-emergent NHSE,  remains robust even under strongly non-ideal anti-symmetric disorder,  extending its applicability beyond fine-tuned scenarios.

\begin{figure}[!tb]
	\centering
	\includegraphics[width=18cm]{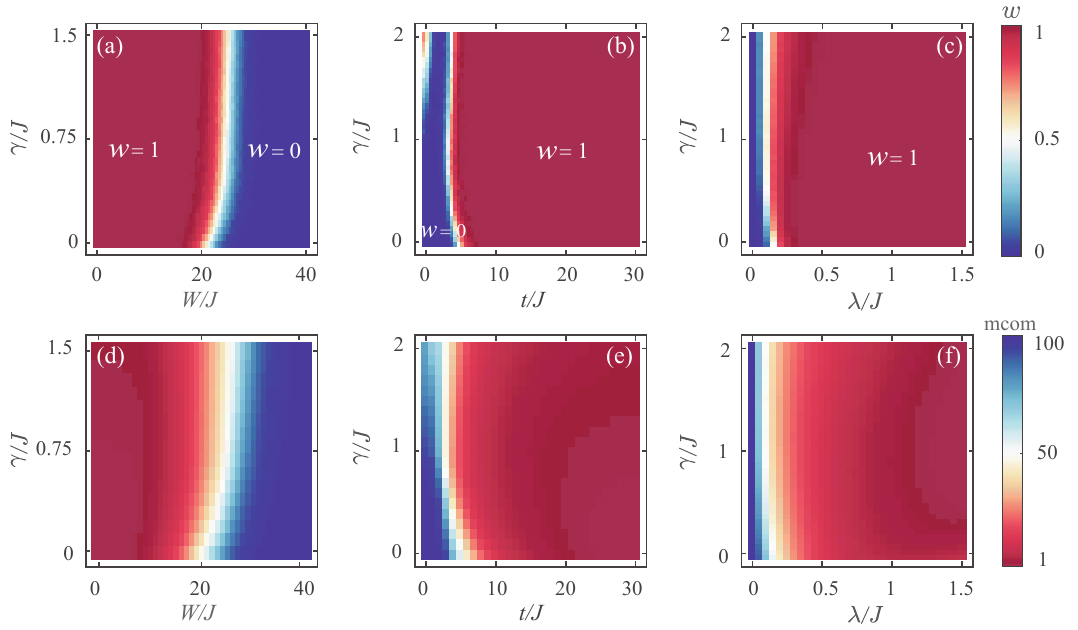}
	\caption{(a-c) Winding number $w$ and (d-f) mcom, under anti-symmetric disorder, in the $(W, \gamma)$ plane with $\lambda/J =1$ and $t/J=14$ (a,d), the $(t, \gamma)$ plane  with $W/J=12$ and $\lambda/J =1$ (b,e), and the $(\lambda, \gamma)$ plane  with  $W/J=12$ and $t/J=14$ (c,f). There exist three regions: Anderson localization with $\textrm{mcom} \sim  N/2$, skin-mode localization with $\textrm{mcom} \sim  1$,  and a mixture of the two with $1 \ll \textrm{mcom} \ll  N/2$, respectively. The results  are averaged over $1200$ disorder realizations with   $N=200$. } \label{FigS8}
\end{figure}

\section{Dependence of   phase diagrams on  $\gamma$}

In the main text, we plot the phase regions of Anderson localization and skin-mode localization as functions of $W$, $\gamma$ and $\lambda$ by fixing $\lambda$.  Now, we consider the dependence of phase diagrams on $\gamma$ under anti-symmetric disorder.

The phase diagrams determined by the winding number in the presence of anti-symmetric disorder are shown in Fig.~\ref{FigS8}(a–c), where the phase boundary between the absence ($w=0$) and presence ($w=1$) of skin modes is clearly visible. However, the winding number cannot identify the coexistence regions of skin-mode localization and Anderson localization. In contrast, this distinction can be made using the mcom, as shown in Fig.~\ref{FigS8}(d–f). Three distinct regions are observed: Anderson localization with $\textrm{mcom} \sim N/2$, skin-mode localization with $\textrm{mcom} \sim 1$, and a mixed phase characterized by $1 \ll \textrm{mcom} \ll N/2$. We find that the phase boundary becomes less sensitive to variations in $\gamma$ once $\gamma$ is sufficiently large. This results from the weakening of nonreciprocal hopping at large values of $\gamma$.

\section{Coexistence region of Anderson and skin-mode localization }

As shown in Fig.~3(d-f) of the main text, under strong disorder, the system exhibits Anderson localization for small inter-chain coupling $t$ (region \Rmnum{1}). As the inter-chain coupling $t$ increases, the system enters region \Rmnum{3}, where Anderson localization and skin-mode localization coexist. This is because increasing $t$ effectively reduces the disorder strength, leading to competition between Anderson localization and the NHSE. With further increase in $t$, the system transitions into region \Rmnum{2}, where all eigenstates exhibit skin-mode localization due to the NHSE. In this region \Rmnum{2}, the effective  disorder   becomes  quite weak at the large inter-chain coupling $t$ for the anti-symmetric disorder configuration.

\begin{figure*}[!tb]
	\centering
	\includegraphics[width=17.5cm]{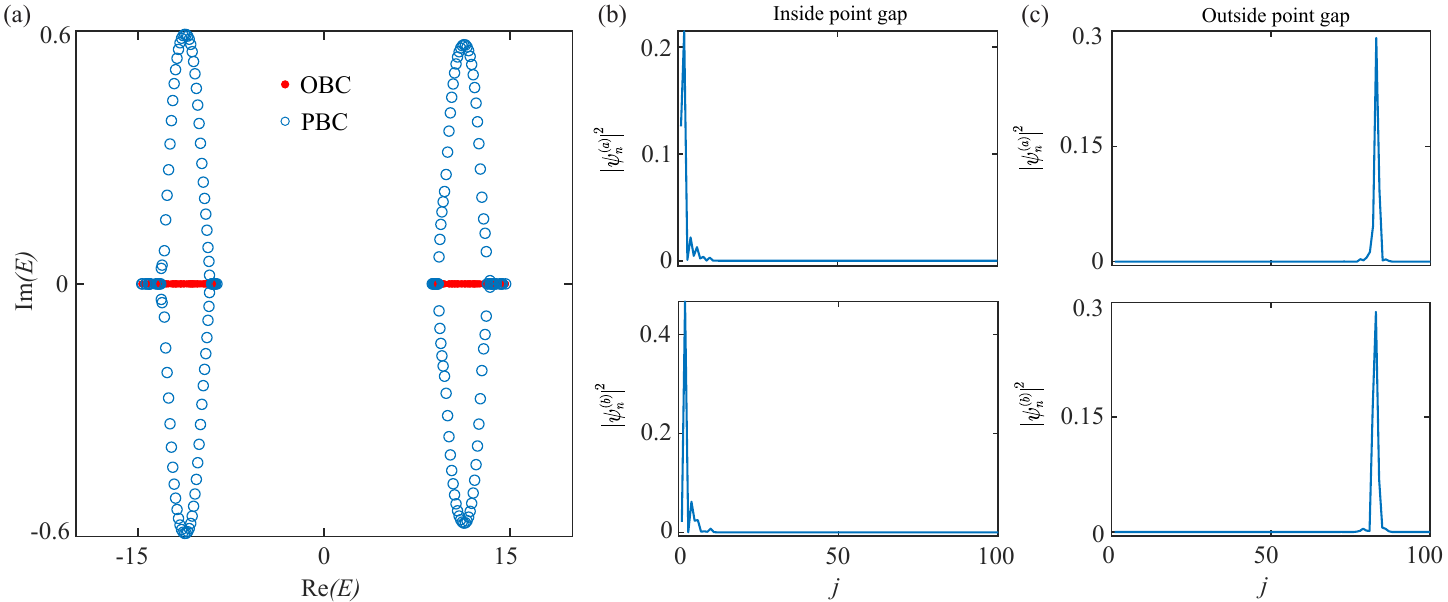}
	\caption{ (a) Complex eigenenergy spectra under OBCs (red dots) and PBCs (blue circles) in region \Rmnum{3}, where the Anderson localization and skin-mode localization coexist. (b,c)  Probability density distributions $|\psi_{n}^{(a)}(j)|^2$ and $|\psi_{n}^{(b)}(j)|^2$ of specific eigenstates inside (b) and outside (c) the point gap.  The   parameters are   $W/J = 20$, $t/J = 10$, and $\gamma/J = \lambda/J = 1$. }\label{FigS10}
\end{figure*}

As explained in the main text, the winding number cannot identify the coexistence region (Region \Rmnum{3}), where both Anderson localization and skin-mode localization are present. Specifically, it cannot distinguish Region \Rmnum{3} from the pure skin-mode localization region (Region \Rmnum{2}). This limitation is expected, as the winding number reflects the presence of the NHSE. Even in the coexistence region, eigenstates under OBCs still exhibit skin-mode localization if their corresponding eigenvalues lie inside the point gap of the PBC spectrum, as illustrated in Fig.~\ref{FigS10}(a). Indeed, part of the OBC spectrum is enclosed by the point gap in the PBC spectrum. Eigenstates associated with eigenvalues inside this point gap are skin-mode localized [see Fig.~\ref{FigS10}(b)], whereas those outside the point gap are Anderson localized [see Fig.~\ref{FigS10}(c)]. Consequently, the winding number remains fixed at $W=1$ throughout Region \Rmnum{3}, resulting in only two distinct regions being visible in the winding number phase diagram.

\begin{figure}[!b]
	\centering
	\includegraphics[width=0.98\textwidth]{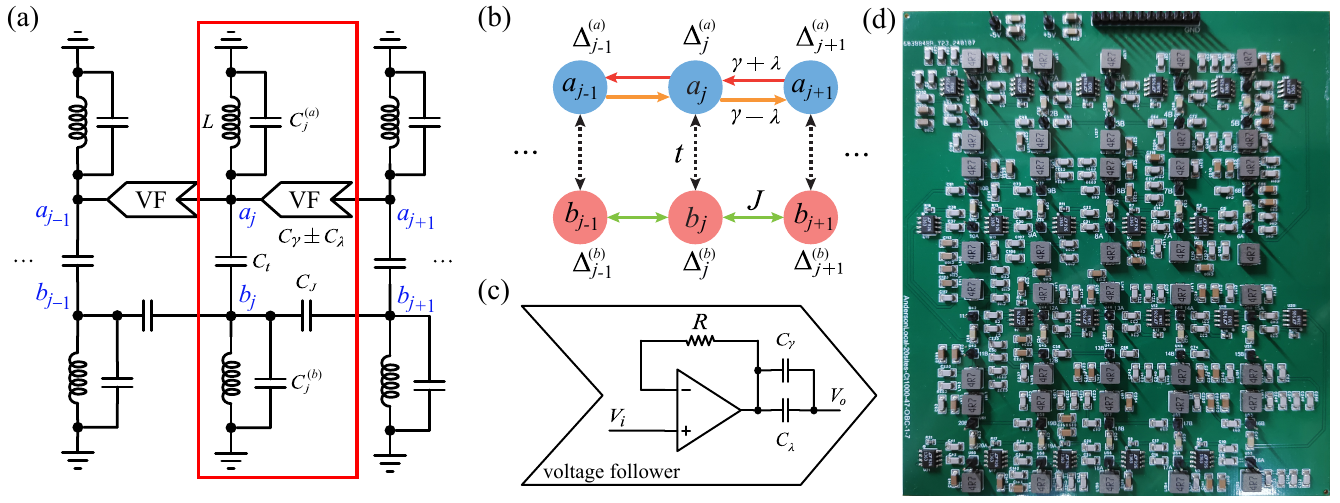}
	\caption{Electrical-circuit lattice. (a) Circuit implementation of the coupled HN-Hermitian lattice subject to   correlated disorder, corresponding to the tight-binding lattice model in (b). The red solid line outlines a unit cell, containing two sublattices $a_j$ and $b_j$, in the circuit lattice. For the HN chain $\{a_j\}$, the nonreciprocal intra-chain hopping, represented by capacitors $C_{\gamma} \pm C_{\lambda}$, is realized by the voltage follower (VF). The circuit diagram of the voltage follower module is shown in (c), where the resistor $R=1~\text{k}\Omega$ in the voltage follower is used to ensure its stability. The capacitor $C_J$ denotes the intra-chain hopping in the Hermitian chain $\{b_j\}$, $C_t$ represents the inter-chain hopping. $C_j^{(a)}$ and $C_j^{(b)}$ are disordered capacitances for simulating the  correlated disorder. The inductor $L$ is used to adjust the   resonance frequency of the circuit.  (d) Photograph of the  whole printed circuit board.}
	\label{figs2}
\end{figure}

\section{Circuit Implementation of the model}

In this section, we present a detailed circuit implementation of our model. Linear circuit networks, composed of linear components, can be characterized by a series of time-dependent differential equations. After applying the Fourier transformation with respect to   time, these equations can be simplified into a set of algebraic equations in the frequency domain. In the frequency domain, the relation of current and voltage between two nodes can be written as
\begin{align}\label{eq1A}
	I_{jk}(\omega) = \frac{V_j(\omega) - V_k(\omega)}{Z_{jk}(\omega)},
\end{align}
where $Z_{jk}(\omega)$ is the impedance between node $j$ and node $k$, and the impedances of capacitor, inductor and resistor are $Z_C(\omega) = 1/i\omega C, Z_L(\omega) = i\omega L$ and $Z_R(\omega)=R$. According to Kirchhoff's current law, the sum of all currents   entering and
leaving a node equals zero. This indicates that the input current $I_j$ at the node $j$ equals the sum of the currents leaving node $j$: 
\begin{align}\label{eq2A}
	I_j = \sum_{k} I_{jk}.
\end{align}

According to Eqs.~(\ref{eq1A}) and  (\ref{eq2A}), we can derive the circuit Laplacian of the electrical circuit  in Fig.~\ref{figs2}(a), corresponding to the tight-binding lattice model in Fig.~\ref{figs2}(b).

The equivalent circuit of the model is composed of inductors and capacitors. The voltage follower is used to equivalently simulate the non-reciprocal hopping of the model with $\gamma=\lambda$. The circuit diagram of the voltage follower module is shown in Fig.~\ref{figs2}(c), where the resistor $R=1~\text{k}\Omega$ in the voltage follower is used to ensure its stability. Capacitors are used to equivalently simulate the reciprocal hopping and onsite potential of the tight-binding model, where the capacitances of the capacitors $C_{\gamma}$, $C_{\lambda}$, $C_t$ and $C_J$ correspond to  the reciprocal hopping strengths   $\gamma$, $\lambda$, $t$ and $J$ in Fig.~\ref{figs2}(b). The correlated disorders $\Delta_j^{(a)}$ and $\Delta_j^{(b)}$ are simulated by the disordered capacitances $C_j^{(a)}$ and $C_j^{(b)}$. In addition, the inductor $L$ is used to adjust the resonance frequency of the circuit. The fabricated circuit boards are shown in Fig.~\ref{figs2}(d).

\begin{figure}[!t]
	\centering
	\includegraphics[width=17cm]{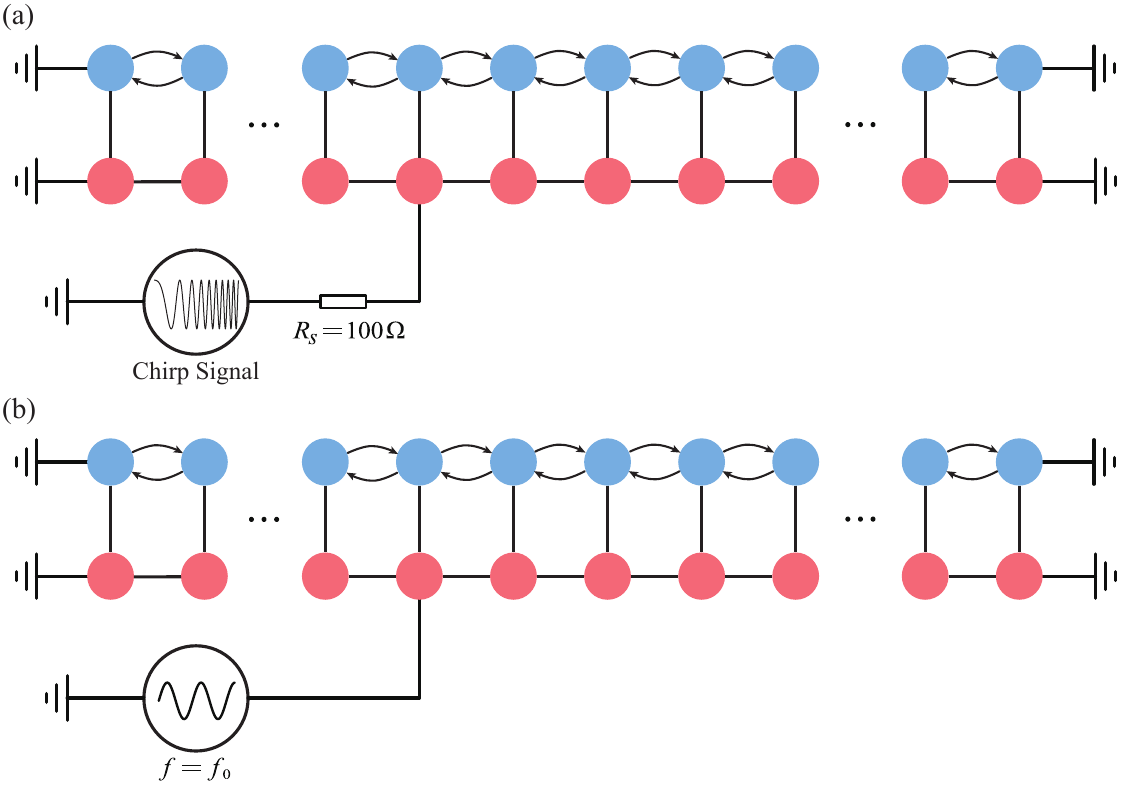}
	\caption{Schematic of the experiment. (a) A chirp signal is used for analyzing the steady-state voltage response at different frequencies. (b) A monochromatic signal  is used for measuring the temporal voltage response of the circuit under the specific frequency $f_0$.}
	\label{figS3}
\end{figure}

The coupled HN-Hermitian chains in the main text can be represented by the   Laplacian $J (\omega)$ of the circuit. The Laplacian is defined as the response of the voltage vector  $\mathbf{V}$ to the input current vector $\mathbf{I}$  by 
\begin{align}\label{eq2}
	\mathbf{I}(\omega) = J (\omega)\mathbf{V}(\omega). 
\end{align}
According to Eq.~(\ref{eq1A}) and Eq.~(\ref{eq2A}),  the Kirchhoff equation of the circuit in Fig.~\ref{figs2}(a) is written as
\begin{align}\label{eq31}
	I_{a,j}= ~&i\omega \left( C_{\gamma}+C_{\lambda} \right) \left( V_{a,j+1}-V_{a,j} \right) +i\omega \left( C_{\gamma}-C_{\lambda} \right) \left( V_{a,j-1}-V_{a,j} \right) +i\omega C_t\left( V_{b,j}-V_{a,j} \right) \nonumber \\
	& -i\omega \left( C_g+C_{j}^{(a)} \right) V_{a,j}-\frac{1}{i\omega L}V_{a,j},
\end{align}
\begin{align}\label{eq32}
	I_{b,j}&=i\omega C_t\left( V_{a,j}-V_{b,j} \right) +i\omega C_J\left( V_{b,j+1}-V_{b,j} \right) +i\omega C_J\left( V_{b,j-1}-V_{b,j} \right)-i\omega \left(C_g+C_{j}^{(b)} \right) V_{b,j}-\frac{1}{i\omega L}V_{b,j},
\end{align}
where $I_{a,j}$ ($V_{a,j}$) and $I_{b,j}$ ($V_{b,j}$) denote the currents (voltages) on $a$ and $b$ sublattices in the $j$th cell, respectively, $\omega$ denotes the circuit frequency, and $C_g$ is grounded capacitor for ensuring the circuit stability. In experiments, we take $C_{\gamma}=C_J$. Then, using Eqs.~(\ref{eq31}) and (\ref{eq32}), the circuit Laplacian $J (\omega)$ is rewritten as 
\begin{align}\label{eq41}
	J(\omega) = i\omega\mathcal{H} _{\textrm{c}}-\left( 2i\omega C_J+i\omega C_t+i\omega C_g+\frac{1}{i\omega L} \right) \mathds{1},
\end{align}
where $\mathds{1}$ is the $2N \times 2N$ identity matrix, and $\mathcal{H} _{\textrm{c}}$ reads
\begin{equation}\label{eq71}
	\mathcal{H}_\textrm{c}= 
	\begingroup
	\setlength{\tabcolsep}{10pt}             
	\renewcommand{\arraystretch}{2.0}        
	\begin{pmatrix}
		- C_{1}^{\left( a \right)}  &		C_t&		C_J+C_{\lambda}&		0&		\cdots&		0&		0\\
		C_t&		- C_{1}^{\left( b \right)}  &		0&		C_J&		\cdots&		0&		0\\
		C_J-C_{\lambda}&		0&		- C_{2}^{\left( a \right)}  &		C_t&		\cdots&		0&		0\\
		0&		C_J&		C_t&		- C_{2}^{\left( b \right)}  &		\cdots&		0&		0\\
		\vdots&		\vdots&		\vdots&		\vdots&		\ddots&		\vdots&		\vdots\\
		0&		0&		0&		0&		\cdots&		- C_{L}^{\left( a \right)}  &		C_t\\
		0&		0&		0&		0&		\cdots&		C_t&		- C_{L}^{\left( b \right)}  \\ 
	\end{pmatrix},
	\endgroup
\end{equation}

The matrix $\mathcal{H} _{\textrm{c}}$ in the first term of the circuit Laplacian $J (\omega)$ in Eq.~(\ref{eq41}) replicates the Hamiltonian matrix of the coupled HN-Hermitian chains described in the main text. The second term of the circuit Laplacian $J (\omega)$ in Eq.~(\ref{eq41}) does not influence the state localization and delocalization. Therefore, this allows the circuit Laplacian to model the desired Hamiltonian.

When the input current is zero, we can obtain the eigenvalue equation:
\begin{align}\label{eq8}
	\mathcal{H} _{\textrm{c}}\mathbf{V}= \left( 2C_J+C_t+C_g-\frac{1}{\omega ^2L} \right) \mathbf{V}.
\end{align}
This indicates that the voltage distribution reflects the state distribution corresponding to the specific eigenvalue by adjusting the frequency $f = \omega/(2\pi)$ of the excited voltage. By measuring the voltage response of the excitation, we can experimentally verify the state delocalization and localization  of the coupled HN-Hermitian chains subject to the correlated disorder.

\begin{figure*}[!t]
	\centering
	\includegraphics[width=18cm]{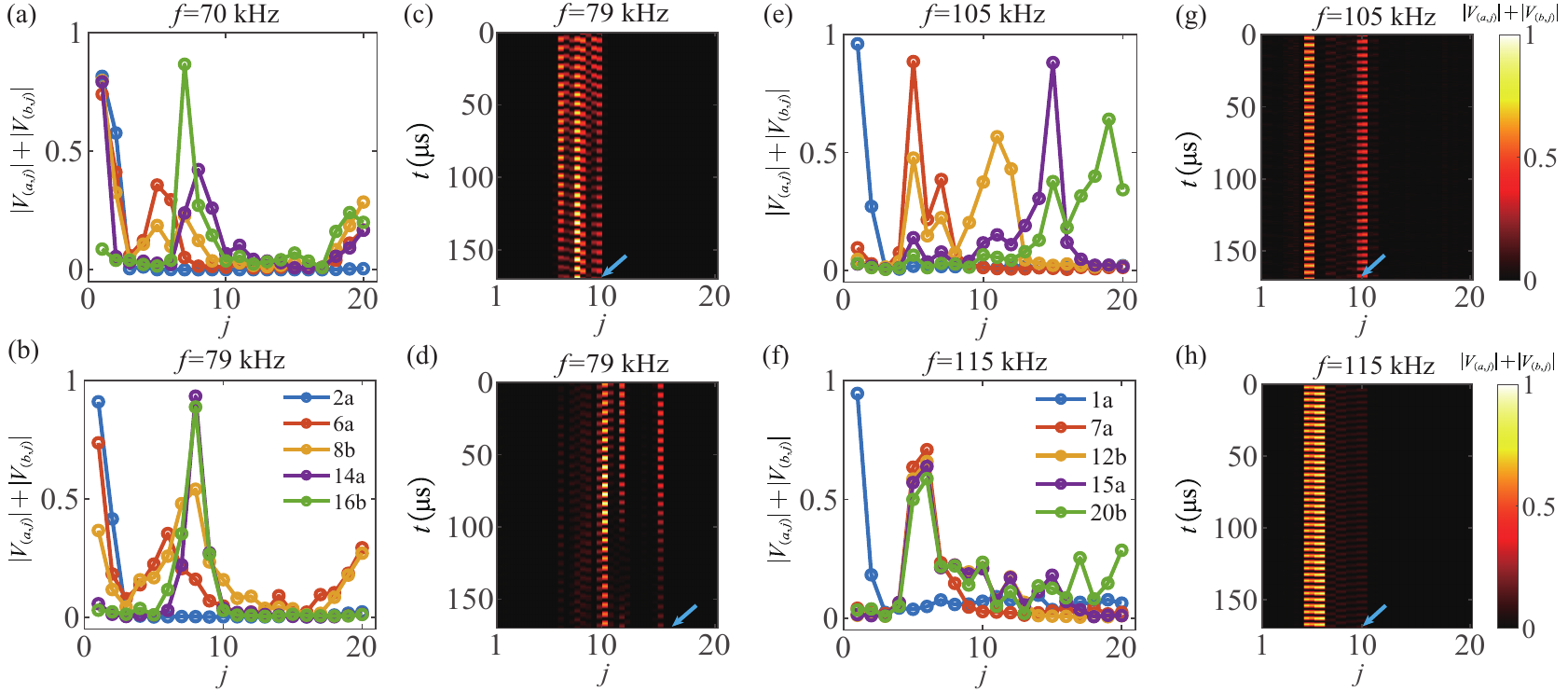}
	\caption{Experimental observations of the Anderson localization of HN-Hermitian coupled chains subject to  symmetric disorder.  (a,b) Measured site-resolved voltage distributions $\abs{V_{a,j}}+\abs{V_{b,j}}$ ($j$ is unit-cell index)  at resonance frequencies $f=70 ~\textrm{kHz}$ (a) and $f=79 ~\textrm{kHz}$ (\textbf{b}) for  weak inter-chain  hopping $C_t$, with $C_{\gamma}=C_{\lambda}=C_J=C_t=100 ~\text{nF}$,   $L=4.7~\mu\text{H}$, and $C_{j}^{\left( a \right)} = C_{j}^{\left( b \right)} \in [-6 C_J,~6 C_J]$. The legend ``$j\alpha$" ($\alpha=a,b$) in (\textbf{b}) indicates the excitation at the $j$th site of the chain $\alpha$. (c,d) Measured temporal voltage responses excited at the $11$th (c) and $17$th (d) unit cells, indicated by the blue arrows,   for the weak inter-chain  hopping with $f=79 ~\textrm{kHz}$.   (e,f) Measured voltage distributions  and (g,h) temporal voltage responses for  strong inter-chain  hopping $C_t$, with $C_{\gamma}=C_{\lambda}=C_J=47~\text{nF}$, $C_t=1~\mu\text{F}$,   $L=4.7~\mu\text{H}$, and $C_{j}^{\left( a \right)} = C_{j}^{\left( b \right)} \in [-6 C_J,~6 C_J]$. } \label{FigS4}
\end{figure*}

A printed circuit board (PCB) layout is simulated using LTSpice, designed and manufactured using LCEDA. The non-reciprocal hopping is implemented using voltage followers based on the unity-gain stable operational amplifier model LT1363. In the circuit, $1~\mathrm{k\Omega}$ ($\pm1\%$ tolerance) resistors are employed to maintain the stability of the operational amplifier, and $100~\mathrm{nF}$ ($\pm10\%$ tolerance) and $2.2~\mu\mathrm{F}$ ($\pm10\%$ tolerance) capacitors are used to suppress ripple noise from the direct current (DC) power supply.

To measure   eigenstates, a signal generator (Keysight 33500B) is used to generate a chirp signal, which is then injected into the circuit. The voltage response to this signal is recorded using an oscilloscope (Keysight DSOX4024A) to 
diagnose the resonance frequency $f$. With this frequency $f$, the circuit is excited at a certain node, and the voltage response is measured on all nodes to obtain the eigenstates. For temporal measurements, the signal generator is utilized to produce a sinusoidal signal with frequency $f$ and duration of $150~\mu\textrm{s}$, and this signal is injected to a certain node of the circuit. The oscilloscope then measures the voltage response at all the nodes to obtain the time-evolving states.

In experiments, we excite the circuit at a certain node, and then test the voltage response at all nodes. The  schematic of the testing is shown in Fig.~\ref{figS3}. A chirp signal (10 $\textrm{kHz}$ - 400 $\textrm{kHz}$) is used for analyzing the steady-state voltage response at  different frequencies [see Fig.~\ref{figS3}(a)]. A monochromatic signal, e.g. sinusoidal wave, is used for measuring the temporal voltage response of the circuit under the specific frequency [see Fig.~\ref{figS3}(b)].

\section{Experimental results of HN-Hermitian coupled chains subject to the symmetric disorder}

As shown in the main text, the experimental results have proved that the strong inter-chain coupling can lead to the re‐emergent NHSE in the coupled NH-Hermitian  chains in spite of the strong anti-symmetric disorder. In this section, as a comparison, we present experimental results of the coupled HN-Hermitian  chains subject to  symmetrically-correlated disorder. We resonantly excite the circuit, and measured the site-resolved  voltage distributions  $\abs{V_{a,j}}+\abs{V_{b,j}}$ in the presence of  symmetric disorder. As shown in Fig.~\ref{FigS4}(a,b) for  weak inter-chain coupling and Fig.~\ref{FigS4}(e,f) for  strong inter-chain coupling,  the voltage distributions remain  mostly localized.  Furthermore, we measure the temporal voltage response, as shown in Fig.~\ref{FigS4}(c,d) for  weak inter-chain coupling and Fig.~\ref{FigS4}(g,h) for  strong inter-chain coupling. Once it is excited, the voltage remains localized as time evolves.


\begin{thebibliography}{68}%
	\makeatletter
	\providecommand \@ifxundefined [1]{%
		\@ifx{#1\undefined}
	}%
	\providecommand \@ifnum [1]{%
		\ifnum #1\expandafter \@firstoftwo
		\else \expandafter \@secondoftwo
		\fi
	}%
	\providecommand \@ifx [1]{%
		\ifx #1\expandafter \@firstoftwo
		\else \expandafter \@secondoftwo
		\fi
	}%
	\providecommand \natexlab [1]{#1}%
	\providecommand \enquote  [1]{``#1''}%
	\providecommand \bibnamefont  [1]{#1}%
	\providecommand \bibfnamefont [1]{#1}%
	\providecommand \citenamefont [1]{#1}%
	\providecommand \href@noop [0]{\@secondoftwo}%
	\providecommand \href [0]{\begingroup \@sanitize@url \@href}%
	\providecommand \@href[1]{\@@startlink{#1}\@@href}%
	\providecommand \@@href[1]{\endgroup#1\@@endlink}%
	\providecommand \@sanitize@url [0]{\catcode `\\12\catcode `\$12\catcode
		`\&12\catcode `\#12\catcode `\^12\catcode `\_12\catcode `\%12\relax}%
	\providecommand \@@startlink[1]{}%
	\providecommand \@@endlink[0]{}%
	\providecommand \url  [0]{\begingroup\@sanitize@url \@url }%
	\providecommand \@url [1]{\endgroup\@href {#1}{\urlprefix }}%
	\providecommand \urlprefix  [0]{URL }%
	\providecommand \Eprint [0]{\href }%
	\providecommand \doibase [0]{http://dx.doi.org/}%
	\providecommand \selectlanguage [0]{\@gobble}%
	\providecommand \bibinfo  [0]{\@secondoftwo}%
	\providecommand \bibfield  [0]{\@secondoftwo}%
	\providecommand \translation [1]{[#1]}%
	\providecommand \BibitemOpen [0]{}%
	\providecommand \bibitemStop [0]{}%
	\providecommand \bibitemNoStop [0]{.\EOS\space}%
	\providecommand \EOS [0]{\spacefactor3000\relax}%
	\providecommand \BibitemShut  [1]{\csname bibitem#1\endcsname}%
	\let\auto@bib@innerbib\@empty
	\bibitem [{\citenamefont {Leykam}\ \emph {et~al.}(2017)\citenamefont {Leykam},
		\citenamefont {Bliokh}, \citenamefont {Huang}, \citenamefont {Chong},\ and\
		\citenamefont {Nori}}]{PhysRevLett.118.040401}%
	\BibitemOpen
	\bibfield  {author} {\bibinfo {author} {\bibfnamefont {D.}~\bibnamefont
			{Leykam}}, \bibinfo {author} {\bibfnamefont {K.~Y.}\ \bibnamefont {Bliokh}},
		\bibinfo {author} {\bibfnamefont {C.}~\bibnamefont {Huang}}, \bibinfo
		{author} {\bibfnamefont {Y.~D.}\ \bibnamefont {Chong}}, \ and\ \bibinfo
		{author} {\bibfnamefont {F.}~\bibnamefont {Nori}},\ }\bibfield  {title}
	{\enquote {\bibinfo {title} {Edge modes, degeneracies, and topological
				numbers in non-\uppercase{H}ermitian systems},}\ }\href
	{https://link.aps.org/doi/10.1103/PhysRevLett.118.040401} {\bibfield
		{journal} {\bibinfo  {journal} {Phys. Rev. Lett.}\ }\textbf {\bibinfo
			{volume} {118}},\ \bibinfo {pages} {040401} (\bibinfo {year}
		{2017})}\BibitemShut {NoStop}%
	\bibitem [{\citenamefont {El-Ganainy}\ \emph {et~al.}(2018)\citenamefont
		{El-Ganainy}, \citenamefont {Makris}, \citenamefont {Khajavikhan},
		\citenamefont {Musslimani}, \citenamefont {Rotter},\ and\ \citenamefont
		{Christodoulides}}]{El-Ganainy2018}%
	\BibitemOpen
	\bibfield  {author} {\bibinfo {author} {\bibfnamefont {R.}~\bibnamefont
			{El-Ganainy}}, \bibinfo {author} {\bibfnamefont {K.~G.}\ \bibnamefont
			{Makris}}, \bibinfo {author} {\bibfnamefont {M.}~\bibnamefont {Khajavikhan}},
		\bibinfo {author} {\bibfnamefont {Z.~H.}\ \bibnamefont {Musslimani}},
		\bibinfo {author} {\bibfnamefont {S.}~\bibnamefont {Rotter}}, \ and\ \bibinfo
		{author} {\bibfnamefont {D.~N.}\ \bibnamefont {Christodoulides}},\ }\bibfield
	{title} {\enquote {\bibinfo {title} {Non-\uppercase{H}ermitian physics and
				\uppercase{PT} symmetry},}\ }\href {http://dx.doi.org/10.1038/nphys4323
		http://10.0.4.14/nphys4323} {\bibfield  {journal} {\bibinfo  {journal} {Nat.
				Phys.}\ }\textbf {\bibinfo {volume} {14}},\ \bibinfo {pages} {11} (\bibinfo
		{year} {2018})}\BibitemShut {NoStop}%
	\bibitem [{\citenamefont {Yao}\ and\ \citenamefont
		{Wang}(2018)}]{ShunyuYao2018}%
	\BibitemOpen
	\bibfield  {author} {\bibinfo {author} {\bibfnamefont {S.}~\bibnamefont
			{Yao}}\ and\ \bibinfo {author} {\bibfnamefont {Z.}~\bibnamefont {Wang}},\
	}\bibfield  {title} {\enquote {\bibinfo {title} {Edge states and topological
				invariants of non-\uppercase{H}ermitian systems},}\ }\href
	{https://link.aps.org/doi/10.1103/PhysRevLett.121.086803} {\bibfield
		{journal} {\bibinfo  {journal} {Phys. Rev. Lett.}\ }\textbf {\bibinfo
			{volume} {121}},\ \bibinfo {pages} {086803} (\bibinfo {year}
		{2018})}\BibitemShut {NoStop}%
	\bibitem [{\citenamefont {Yokomizo}\ and\ \citenamefont
		{Murakami}(2019)}]{PhysRevLett.123.066404}%
	\BibitemOpen
	\bibfield  {author} {\bibinfo {author} {\bibfnamefont {K.}~\bibnamefont
			{Yokomizo}}\ and\ \bibinfo {author} {\bibfnamefont {S.}~\bibnamefont
			{Murakami}},\ }\bibfield  {title} {\enquote {\bibinfo {title} {Non-{B}loch
				band theory of non-{H}ermitian systems},}\ }\href {\doibase
		10.1103/PhysRevLett.123.066404} {\bibfield  {journal} {\bibinfo  {journal}
			{Phys. Rev. Lett.}\ }\textbf {\bibinfo {volume} {123}},\ \bibinfo {pages}
		{066404} (\bibinfo {year} {2019})}\BibitemShut {NoStop}%
	\bibitem [{\citenamefont {Liu}\ \emph {et~al.}(2019)\citenamefont {Liu},
		\citenamefont {Zhang}, \citenamefont {Ai}, \citenamefont {Gong},
		\citenamefont {Kawabata}, \citenamefont {Ueda},\ and\ \citenamefont
		{Nori}}]{PhysRevLett.122.076801}%
	\BibitemOpen
	\bibfield  {author} {\bibinfo {author} {\bibfnamefont {T.}~\bibnamefont
			{Liu}}, \bibinfo {author} {\bibfnamefont {Y.-R.}\ \bibnamefont {Zhang}},
		\bibinfo {author} {\bibfnamefont {Q.}~\bibnamefont {Ai}}, \bibinfo {author}
		{\bibfnamefont {Z.}~\bibnamefont {Gong}}, \bibinfo {author} {\bibfnamefont
			{K.}~\bibnamefont {Kawabata}}, \bibinfo {author} {\bibfnamefont
			{M.}~\bibnamefont {Ueda}}, \ and\ \bibinfo {author} {\bibfnamefont
			{F.}~\bibnamefont {Nori}},\ }\bibfield  {title} {\enquote {\bibinfo {title}
			{Second-order topological phases in non-{H}ermitian systems},}\ }\href
	{\doibase 10.1103/PhysRevLett.122.076801} {\bibfield  {journal} {\bibinfo
			{journal} {Phys. Rev. Lett.}\ }\textbf {\bibinfo {volume} {122}},\ \bibinfo
		{pages} {076801} (\bibinfo {year} {2019})}\BibitemShut {NoStop}%
	\bibitem [{\citenamefont {Zhang}\ \emph {et~al.}(2020)\citenamefont {Zhang},
		\citenamefont {Yang},\ and\ \citenamefont {Fang}}]{PhysRevLett.125.126402}%
	\BibitemOpen
	\bibfield  {author} {\bibinfo {author} {\bibfnamefont {K.}~\bibnamefont
			{Zhang}}, \bibinfo {author} {\bibfnamefont {Z.}~\bibnamefont {Yang}}, \ and\
		\bibinfo {author} {\bibfnamefont {C.}~\bibnamefont {Fang}},\ }\bibfield
	{title} {\enquote {\bibinfo {title} {Correspondence between winding numbers
				and skin modes in non-{H}ermitian systems},}\ }\href {\doibase
		10.1103/PhysRevLett.125.126402} {\bibfield  {journal} {\bibinfo  {journal}
			{Phys. Rev. Lett.}\ }\textbf {\bibinfo {volume} {125}},\ \bibinfo {pages}
		{126402} (\bibinfo {year} {2020})}\BibitemShut {NoStop}%
	\bibitem [{\citenamefont {Okuma}\ \emph {et~al.}(2020)\citenamefont {Okuma},
		\citenamefont {Kawabata}, \citenamefont {Shiozaki},\ and\ \citenamefont
		{Sato}}]{PhysRevLett.124.086801}%
	\BibitemOpen
	\bibfield  {author} {\bibinfo {author} {\bibfnamefont {N.}~\bibnamefont
			{Okuma}}, \bibinfo {author} {\bibfnamefont {K.}~\bibnamefont {Kawabata}},
		\bibinfo {author} {\bibfnamefont {K.}~\bibnamefont {Shiozaki}}, \ and\
		\bibinfo {author} {\bibfnamefont {M.}~\bibnamefont {Sato}},\ }\bibfield
	{title} {\enquote {\bibinfo {title} {Topological origin of non-{H}ermitian
				skin effects},}\ }\href {\doibase 10.1103/PhysRevLett.124.086801} {\bibfield
		{journal} {\bibinfo  {journal} {Phys. Rev. Lett.}\ }\textbf {\bibinfo
			{volume} {124}},\ \bibinfo {pages} {086801} (\bibinfo {year}
		{2020})}\BibitemShut {NoStop}%
	\bibitem [{\citenamefont {Kawabata}\ \emph
		{et~al.}(2019{\natexlab{a}})\citenamefont {Kawabata}, \citenamefont
		{Bessho},\ and\ \citenamefont {Sato}}]{PhysRevLett.123.066405}%
	\BibitemOpen
	\bibfield  {author} {\bibinfo {author} {\bibfnamefont {K.}~\bibnamefont
			{Kawabata}}, \bibinfo {author} {\bibfnamefont {T.}~\bibnamefont {Bessho}}, \
		and\ \bibinfo {author} {\bibfnamefont {M.}~\bibnamefont {Sato}},\ }\bibfield
	{title} {\enquote {\bibinfo {title} {Classification of exceptional points and
				non-{H}ermitian topological semimetals},}\ }\href {\doibase
		10.1103/PhysRevLett.123.066405} {\bibfield  {journal} {\bibinfo  {journal}
			{Phys. Rev. Lett.}\ }\textbf {\bibinfo {volume} {123}},\ \bibinfo {pages}
		{066405} (\bibinfo {year} {2019}{\natexlab{a}})}\BibitemShut {NoStop}%
	\bibitem [{\citenamefont {Ge}\ \emph {et~al.}(2019)\citenamefont {Ge},
		\citenamefont {Zhang}, \citenamefont {Liu}, \citenamefont {Li}, \citenamefont
		{Fan},\ and\ \citenamefont {Nori}}]{PhysRevB.100.054105}%
	\BibitemOpen
	\bibfield  {author} {\bibinfo {author} {\bibfnamefont {Z.~Y.}\ \bibnamefont
			{Ge}}, \bibinfo {author} {\bibfnamefont {Y.~R.}\ \bibnamefont {Zhang}},
		\bibinfo {author} {\bibfnamefont {T.}~\bibnamefont {Liu}}, \bibinfo {author}
		{\bibfnamefont {S.~W.}\ \bibnamefont {Li}}, \bibinfo {author} {\bibfnamefont
			{H.}~\bibnamefont {Fan}}, \ and\ \bibinfo {author} {\bibfnamefont
			{F.}~\bibnamefont {Nori}},\ }\bibfield  {title} {\enquote {\bibinfo {title}
			{Topological band theory for non-{H}ermitian systems from the {D}irac
				equation},}\ }\href {\doibase 10.1103/PhysRevB.100.054105} {\bibfield
		{journal} {\bibinfo  {journal} {Phys. Rev. B}\ }\textbf {\bibinfo {volume}
			{100}},\ \bibinfo {pages} {054105} (\bibinfo {year} {2019})}\BibitemShut
	{NoStop}%
	\bibitem [{\citenamefont {Zhao}\ \emph {et~al.}(2019)\citenamefont {Zhao},
		\citenamefont {Qiao}, \citenamefont {Wu}, \citenamefont {Midya},
		\citenamefont {Longhi},\ and\ \citenamefont {Feng}}]{Zhao2019}%
	\BibitemOpen
	\bibfield  {author} {\bibinfo {author} {\bibfnamefont {H.}~\bibnamefont
			{Zhao}}, \bibinfo {author} {\bibfnamefont {X.}~\bibnamefont {Qiao}}, \bibinfo
		{author} {\bibfnamefont {T.}~\bibnamefont {Wu}}, \bibinfo {author}
		{\bibfnamefont {B.}~\bibnamefont {Midya}}, \bibinfo {author} {\bibfnamefont
			{S.}~\bibnamefont {Longhi}}, \ and\ \bibinfo {author} {\bibfnamefont
			{L.}~\bibnamefont {Feng}},\ }\bibfield  {title} {\enquote {\bibinfo {title}
			{Non-{H}ermitian topological light steering},}\ }\href {\doibase
		10.1126/science.aay1064} {\bibfield  {journal} {\bibinfo  {journal}
			{Science}\ }\textbf {\bibinfo {volume} {365}},\ \bibinfo {pages} {1163}
		(\bibinfo {year} {2019})}\BibitemShut {NoStop}%
	\bibitem [{\citenamefont {Kawabata}\ \emph
		{et~al.}(2019{\natexlab{b}})\citenamefont {Kawabata}, \citenamefont
		{Shiozaki}, \citenamefont {Ueda},\ and\ \citenamefont
		{Sato}}]{PhysRevX.9.041015}%
	\BibitemOpen
	\bibfield  {author} {\bibinfo {author} {\bibfnamefont {K.}~\bibnamefont
			{Kawabata}}, \bibinfo {author} {\bibfnamefont {K.}~\bibnamefont {Shiozaki}},
		\bibinfo {author} {\bibfnamefont {M.}~\bibnamefont {Ueda}}, \ and\ \bibinfo
		{author} {\bibfnamefont {M.}~\bibnamefont {Sato}},\ }\bibfield  {title}
	{\enquote {\bibinfo {title} {Symmetry and topology in non-{H}ermitian
				physics},}\ }\href {\doibase 10.1103/PhysRevX.9.041015} {\bibfield  {journal}
		{\bibinfo  {journal} {Phys. Rev. X}\ }\textbf {\bibinfo {volume} {9}},\
		\bibinfo {pages} {041015} (\bibinfo {year} {2019}{\natexlab{b}})}\BibitemShut
	{NoStop}%
	\bibitem [{\citenamefont {Borgnia}\ \emph {et~al.}(2020)\citenamefont
		{Borgnia}, \citenamefont {Kruchkov},\ and\ \citenamefont
		{Slager}}]{PhysRevLett.124.056802}%
	\BibitemOpen
	\bibfield  {author} {\bibinfo {author} {\bibfnamefont {D.~S.}\ \bibnamefont
			{Borgnia}}, \bibinfo {author} {\bibfnamefont {A.~J.}\ \bibnamefont
			{Kruchkov}}, \ and\ \bibinfo {author} {\bibfnamefont {R.-J.}\ \bibnamefont
			{Slager}},\ }\bibfield  {title} {\enquote {\bibinfo {title} {Non-{H}ermitian
				boundary modes and topology},}\ }\href {\doibase
		10.1103/PhysRevLett.124.056802} {\bibfield  {journal} {\bibinfo  {journal}
			{Phys. Rev. Lett.}\ }\textbf {\bibinfo {volume} {124}},\ \bibinfo {pages}
		{056802} (\bibinfo {year} {2020})}\BibitemShut {NoStop}%
	\bibitem [{\citenamefont {Liu}\ \emph {et~al.}(2020)\citenamefont {Liu},
		\citenamefont {He}, \citenamefont {Yoshida}, \citenamefont {Xiang},\ and\
		\citenamefont {Nori}}]{PhysRevB.102.235151}%
	\BibitemOpen
	\bibfield  {author} {\bibinfo {author} {\bibfnamefont {T.}~\bibnamefont
			{Liu}}, \bibinfo {author} {\bibfnamefont {J.~J.}\ \bibnamefont {He}},
		\bibinfo {author} {\bibfnamefont {T.}~\bibnamefont {Yoshida}}, \bibinfo
		{author} {\bibfnamefont {Z.-L.}\ \bibnamefont {Xiang}}, \ and\ \bibinfo
		{author} {\bibfnamefont {F.}~\bibnamefont {Nori}},\ }\bibfield  {title}
	{\enquote {\bibinfo {title} {Non-{H}ermitian topological {M}ott insulators in
				one-dimensional fermionic superlattices},}\ }\href {\doibase
		10.1103/PhysRevB.102.235151} {\bibfield  {journal} {\bibinfo  {journal}
			{Phys. Rev. B}\ }\textbf {\bibinfo {volume} {102}},\ \bibinfo {pages}
		{235151} (\bibinfo {year} {2020})}\BibitemShut {NoStop}%
	\bibitem [{\citenamefont {Liu}\ \emph {et~al.}(2021)\citenamefont {Liu},
		\citenamefont {He}, \citenamefont {Yang},\ and\ \citenamefont
		{Nori}}]{PhysRevLett.127.196801}%
	\BibitemOpen
	\bibfield  {author} {\bibinfo {author} {\bibfnamefont {T.}~\bibnamefont
			{Liu}}, \bibinfo {author} {\bibfnamefont {J.~J.}\ \bibnamefont {He}},
		\bibinfo {author} {\bibfnamefont {Z.}~\bibnamefont {Yang}}, \ and\ \bibinfo
		{author} {\bibfnamefont {F.}~\bibnamefont {Nori}},\ }\bibfield  {title}
	{\enquote {\bibinfo {title} {Higher-order {W}eyl-exceptional-ring
				semimetals},}\ }\href {\doibase 10.1103/PhysRevLett.127.196801} {\bibfield
		{journal} {\bibinfo  {journal} {Phys. Rev. Lett.}\ }\textbf {\bibinfo
			{volume} {127}},\ \bibinfo {pages} {196801} (\bibinfo {year}
		{2021})}\BibitemShut {NoStop}%
	\bibitem [{\citenamefont {Mu}\ \emph {et~al.}(2022)\citenamefont {Mu},
		\citenamefont {Zhou}, \citenamefont {Li},\ and\ \citenamefont
		{Gong}}]{PhysRevB.105.205402}%
	\BibitemOpen
	\bibfield  {author} {\bibinfo {author} {\bibfnamefont {S.}~\bibnamefont
			{Mu}}, \bibinfo {author} {\bibfnamefont {L.}~\bibnamefont {Zhou}}, \bibinfo
		{author} {\bibfnamefont {L.}~\bibnamefont {Li}}, \ and\ \bibinfo {author}
		{\bibfnamefont {J.}~\bibnamefont {Gong}},\ }\bibfield  {title} {\enquote
		{\bibinfo {title} {Non-{H}ermitian pseudo mobility edge in a coupled chain
				system},}\ }\href {\doibase 10.1103/PhysRevB.105.205402} {\bibfield
		{journal} {\bibinfo  {journal} {Phys. Rev. B}\ }\textbf {\bibinfo {volume}
			{105}},\ \bibinfo {pages} {205402} (\bibinfo {year} {2022})}\BibitemShut
	{NoStop}%
	\bibitem [{\citenamefont {Li}\ and\ \citenamefont
		{Xu}(2022)}]{PhysRevLett.129.093001}%
	\BibitemOpen
	\bibfield  {author} {\bibinfo {author} {\bibfnamefont {K.}~\bibnamefont
			{Li}}\ and\ \bibinfo {author} {\bibfnamefont {Y.}~\bibnamefont {Xu}},\
	}\bibfield  {title} {\enquote {\bibinfo {title} {Non-{H}ermitian absorption
				spectroscopy},}\ }\href {\doibase 10.1103/PhysRevLett.129.093001} {\bibfield
		{journal} {\bibinfo  {journal} {Phys. Rev. Lett.}\ }\textbf {\bibinfo
			{volume} {129}},\ \bibinfo {pages} {093001} (\bibinfo {year}
		{2022})}\BibitemShut {NoStop}%
	\bibitem [{\citenamefont {Ren}\ \emph {et~al.}(2022)\citenamefont {Ren},
		\citenamefont {Liu}, \citenamefont {Zhao}, \citenamefont {He}, \citenamefont
		{Pak}, \citenamefont {Li},\ and\ \citenamefont {Jo}}]{Ren2022}%
	\BibitemOpen
	\bibfield  {author} {\bibinfo {author} {\bibfnamefont {Z.}~\bibnamefont
			{Ren}}, \bibinfo {author} {\bibfnamefont {D.}~\bibnamefont {Liu}}, \bibinfo
		{author} {\bibfnamefont {E.}~\bibnamefont {Zhao}}, \bibinfo {author}
		{\bibfnamefont {C.}~\bibnamefont {He}}, \bibinfo {author} {\bibfnamefont
			{K.~K.}\ \bibnamefont {Pak}}, \bibinfo {author} {\bibfnamefont
			{J.}~\bibnamefont {Li}}, \ and\ \bibinfo {author} {\bibfnamefont {G.-B.}\
			\bibnamefont {Jo}},\ }\bibfield  {title} {\enquote {\bibinfo {title} {Chiral
				control of quantum states in non-{H}ermitian spin{\textendash}orbit-coupled
				fermions},}\ }\href {\doibase 10.1038/s41567-021-01491-x} {\bibfield
		{journal} {\bibinfo  {journal} {Nat. Phys.}\ }\textbf {\bibinfo {volume}
			{18}},\ \bibinfo {pages} {385} (\bibinfo {year} {2022})}\BibitemShut
	{NoStop}%
	\bibitem [{\citenamefont {Cai}\ \emph {et~al.}(2024{\natexlab{a}})\citenamefont
		{Cai}, \citenamefont {Liu},\ and\ \citenamefont
		{Yang}}]{PhysRevA.109.063329}%
	\BibitemOpen
	\bibfield  {author} {\bibinfo {author} {\bibfnamefont {Z.-F.}\ \bibnamefont
			{Cai}}, \bibinfo {author} {\bibfnamefont {T.}~\bibnamefont {Liu}}, \ and\
		\bibinfo {author} {\bibfnamefont {Z.}~\bibnamefont {Yang}},\ }\bibfield
	{title} {\enquote {\bibinfo {title} {Non-{H}ermitian skin effect in
				periodically driven dissipative ultracold atoms},}\ }\href {\doibase
		10.1103/PhysRevA.109.063329} {\bibfield  {journal} {\bibinfo  {journal}
			{Phys. Rev. A}\ }\textbf {\bibinfo {volume} {109}},\ \bibinfo {pages}
		{063329} (\bibinfo {year} {2024}{\natexlab{a}})}\BibitemShut {NoStop}%
	\bibitem [{\citenamefont {Li}\ \emph {et~al.}(2022)\citenamefont {Li},
		\citenamefont {Teo}, \citenamefont {Mu},\ and\ \citenamefont
		{Gong}}]{PhysRevB.106.085427}%
	\BibitemOpen
	\bibfield  {author} {\bibinfo {author} {\bibfnamefont {L.}~\bibnamefont
			{Li}}, \bibinfo {author} {\bibfnamefont {W.~X.}\ \bibnamefont {Teo}},
		\bibinfo {author} {\bibfnamefont {S.}~\bibnamefont {Mu}}, \ and\ \bibinfo
		{author} {\bibfnamefont {J.}~\bibnamefont {Gong}},\ }\bibfield  {title}
	{\enquote {\bibinfo {title} {Direction reversal of non-{H}ermitian skin
				effect via coherent coupling},}\ }\href {\doibase
		10.1103/PhysRevB.106.085427} {\bibfield  {journal} {\bibinfo  {journal}
			{Phys. Rev. B}\ }\textbf {\bibinfo {volume} {106}},\ \bibinfo {pages}
		{085427} (\bibinfo {year} {2022})}\BibitemShut {NoStop}%
	\bibitem [{\citenamefont {Li}\ \emph {et~al.}(2024)\citenamefont {Li},
		\citenamefont {Cai}, \citenamefont {Liu},\ and\ \citenamefont
		{Nori}}]{arXiv:2408.12451}%
	\BibitemOpen
	\bibfield  {author} {\bibinfo {author} {\bibfnamefont {Y.}~\bibnamefont
			{Li}}, \bibinfo {author} {\bibfnamefont {Z.-F.}\ \bibnamefont {Cai}},
		\bibinfo {author} {\bibfnamefont {T.}~\bibnamefont {Liu}}, \ and\ \bibinfo
		{author} {\bibfnamefont {F.}~\bibnamefont {Nori}},\ }\bibfield  {title}
	{\enquote {\bibinfo {title} {Dissipation and interaction-controlled
				non-{H}ermitian skin effects},}\ }\href
	{https://doi.org/10.48550/arXiv.2408.12451} {\bibfield  {journal} {\bibinfo
			{journal} {arXiv:2408.12451}\ } (\bibinfo {year} {2024})}\BibitemShut
	{NoStop}%
	\bibitem [{\citenamefont {Kawabata}\ \emph {et~al.}(2023)\citenamefont
		{Kawabata}, \citenamefont {Numasawa},\ and\ \citenamefont
		{Ryu}}]{PhysRevX.13.021007}%
	\BibitemOpen
	\bibfield  {author} {\bibinfo {author} {\bibfnamefont {K.}~\bibnamefont
			{Kawabata}}, \bibinfo {author} {\bibfnamefont {T.}~\bibnamefont {Numasawa}},
		\ and\ \bibinfo {author} {\bibfnamefont {S.}~\bibnamefont {Ryu}},\ }\bibfield
	{title} {\enquote {\bibinfo {title} {Entanglement phase transition induced
				by the non-{H}ermitian skin effect},}\ }\href {\doibase
		10.1103/PhysRevX.13.021007} {\bibfield  {journal} {\bibinfo  {journal} {Phys.
				Rev. X}\ }\textbf {\bibinfo {volume} {13}},\ \bibinfo {pages} {021007}
		(\bibinfo {year} {2023})}\BibitemShut {NoStop}%
	\bibitem [{\citenamefont {Ling}\ \emph {et~al.}(2025)\citenamefont {Ling},
		\citenamefont {Cai},\ and\ \citenamefont {Liu}}]{PhysRevB.111.205418}%
	\BibitemOpen
	\bibfield  {author} {\bibinfo {author} {\bibfnamefont {W.-Z.}\ \bibnamefont
			{Ling}}, \bibinfo {author} {\bibfnamefont {Z.-F.}\ \bibnamefont {Cai}}, \
		and\ \bibinfo {author} {\bibfnamefont {T.}~\bibnamefont {Liu}},\ }\bibfield
	{title} {\enquote {\bibinfo {title} {Interaction-induced second-order skin
				effect},}\ }\href {\doibase 10.1103/PhysRevB.111.205418} {\bibfield
		{journal} {\bibinfo  {journal} {Phys. Rev. B}\ }\textbf {\bibinfo {volume}
			{111}},\ \bibinfo {pages} {205418} (\bibinfo {year} {2025})}\BibitemShut
	{NoStop}%
	\bibitem [{\citenamefont {Cai}\ \emph {et~al.}(2025)\citenamefont {Cai},
		\citenamefont {Wang}, \citenamefont {Liang}, \citenamefont {Liu},\ and\
		\citenamefont {Nori}}]{PhysRevAL061701}%
	\BibitemOpen
	\bibfield  {author} {\bibinfo {author} {\bibfnamefont {Z.-F.}\ \bibnamefont
			{Cai}}, \bibinfo {author} {\bibfnamefont {X.}~\bibnamefont {Wang}}, \bibinfo
		{author} {\bibfnamefont {Z.-X.}\ \bibnamefont {Liang}}, \bibinfo {author}
		{\bibfnamefont {T.}~\bibnamefont {Liu}}, \ and\ \bibinfo {author}
		{\bibfnamefont {F.}~\bibnamefont {Nori}},\ }\bibfield  {title} {\enquote
		{\bibinfo {title} {Chiral-extended photon-emitter dressed states in
				non-{H}ermitian topological baths},}\ }\href {\doibase 10.1103/8qpx-68x6}
	{\bibfield  {journal} {\bibinfo  {journal} {Phys. Rev. A}\ }\textbf {\bibinfo
			{volume} {111}},\ \bibinfo {pages} {L061701} (\bibinfo {year}
		{2025})}\BibitemShut {NoStop}%
	\bibitem [{\citenamefont {Cai}\ \emph {et~al.}(2024{\natexlab{b}})\citenamefont
		{Cai}, \citenamefont {Wang}, \citenamefont {Zhang}, \citenamefont {Liu},\
		and\ \citenamefont {Nori}}]{arXiv:2411.10398}%
	\BibitemOpen
	\bibfield  {author} {\bibinfo {author} {\bibfnamefont {Z.-F.}\ \bibnamefont
			{Cai}}, \bibinfo {author} {\bibfnamefont {Y.-C.}\ \bibnamefont {Wang}},
		\bibinfo {author} {\bibfnamefont {Y.-R.}\ \bibnamefont {Zhang}}, \bibinfo
		{author} {\bibfnamefont {T.}~\bibnamefont {Liu}}, \ and\ \bibinfo {author}
		{\bibfnamefont {F.}~\bibnamefont {Nori}},\ }\bibfield  {title} {\enquote
		{\bibinfo {title} {Versatile control of nonlinear topological states in
				non-{H}ermitian systems},}\ }\href@noop {} {\bibfield  {journal} {\bibinfo
			{journal} {arXiv:2411.10398}\ } (\bibinfo {year}
		{2024}{\natexlab{b}})}\BibitemShut {NoStop}%
	\bibitem [{\citenamefont {Regensburger}\ \emph {et~al.}(2012)\citenamefont
		{Regensburger}, \citenamefont {Bersch}, \citenamefont {Miri}, \citenamefont
		{Onishchukov}, \citenamefont {Christodoulides},\ and\ \citenamefont
		{Peschel}}]{Regensburger2012}%
	\BibitemOpen
	\bibfield  {author} {\bibinfo {author} {\bibfnamefont {A.}~\bibnamefont
			{Regensburger}}, \bibinfo {author} {\bibfnamefont {C.}~\bibnamefont
			{Bersch}}, \bibinfo {author} {\bibfnamefont {M.~A.}\ \bibnamefont {Miri}},
		\bibinfo {author} {\bibfnamefont {G.}~\bibnamefont {Onishchukov}}, \bibinfo
		{author} {\bibfnamefont {D.~N.}\ \bibnamefont {Christodoulides}}, \ and\
		\bibinfo {author} {\bibfnamefont {U.}~\bibnamefont {Peschel}},\ }\bibfield
	{title} {\enquote {\bibinfo {title} {Parity–time synthetic photonic
				lattices},}\ }\href {http://dx.doi.org/10.1038/nature11298} {\bibfield
		{journal} {\bibinfo  {journal} {Nature}\ }\textbf {\bibinfo {volume} {488}},\
		\bibinfo {pages} {167} (\bibinfo {year} {2012})}\BibitemShut {NoStop}%
	\bibitem [{\citenamefont {Jing}\ \emph {et~al.}(2014)\citenamefont {Jing},
		\citenamefont {\"Ozdemir}, \citenamefont {L\"u}, \citenamefont {Zhang},
		\citenamefont {Yang},\ and\ \citenamefont {Nori}}]{PhysRevLett.113.053604}%
	\BibitemOpen
	\bibfield  {author} {\bibinfo {author} {\bibfnamefont {H.}~\bibnamefont
			{Jing}}, \bibinfo {author} {\bibfnamefont {S.~K.}\ \bibnamefont {\"Ozdemir}},
		\bibinfo {author} {\bibfnamefont {X.~Y.}\ \bibnamefont {L\"u}}, \bibinfo
		{author} {\bibfnamefont {J.}~\bibnamefont {Zhang}}, \bibinfo {author}
		{\bibfnamefont {L.}~\bibnamefont {Yang}}, \ and\ \bibinfo {author}
		{\bibfnamefont {F.}~\bibnamefont {Nori}},\ }\bibfield  {title} {\enquote
		{\bibinfo {title} {$\mathcal{PT}$-symmetric phonon laser},}\ }\href
	{https://link.aps.org/doi/10.1103/PhysRevLett.113.053604} {\bibfield
		{journal} {\bibinfo  {journal} {Phys. Rev. Lett.}\ }\textbf {\bibinfo
			{volume} {113}},\ \bibinfo {pages} {053604} (\bibinfo {year}
		{2014})}\BibitemShut {NoStop}%
	\bibitem [{\citenamefont {Hodaei}\ \emph {et~al.}(2014)\citenamefont {Hodaei},
		\citenamefont {Miri}, \citenamefont {Heinrich}, \citenamefont
		{Christodoulides},\ and\ \citenamefont {Khajavikhan}}]{Hodaei975}%
	\BibitemOpen
	\bibfield  {author} {\bibinfo {author} {\bibfnamefont {H.}~\bibnamefont
			{Hodaei}}, \bibinfo {author} {\bibfnamefont {M.~A.}\ \bibnamefont {Miri}},
		\bibinfo {author} {\bibfnamefont {M.}~\bibnamefont {Heinrich}}, \bibinfo
		{author} {\bibfnamefont {D.~N.}\ \bibnamefont {Christodoulides}}, \ and\
		\bibinfo {author} {\bibfnamefont {M.}~\bibnamefont {Khajavikhan}},\
	}\bibfield  {title} {\enquote {\bibinfo {title}
			{Parity-time{\textendash}symmetric microring lasers},}\ }\href
	{http://science.sciencemag.org/content/346/6212/975} {\bibfield  {journal}
		{\bibinfo  {journal} {Science}\ }\textbf {\bibinfo {volume} {346}},\ \bibinfo
		{pages} {975} (\bibinfo {year} {2014})}\BibitemShut {NoStop}%
	\bibitem [{\citenamefont {Peng}\ \emph
		{et~al.}(2014{\natexlab{a}})\citenamefont {Peng}, \citenamefont
		{{\"{O}}zdemir}, \citenamefont {Lei}, \citenamefont {Monifi}, \citenamefont
		{Gianfreda}, \citenamefont {Long}, \citenamefont {Fan}, \citenamefont {Nori},
		\citenamefont {Bender},\ and\ \citenamefont {Yang}}]{Peng2014}%
	\BibitemOpen
	\bibfield  {author} {\bibinfo {author} {\bibfnamefont {B.}~\bibnamefont
			{Peng}}, \bibinfo {author} {\bibfnamefont {Ş.~K.}\ \bibnamefont
			{{\"{O}}zdemir}}, \bibinfo {author} {\bibfnamefont {F.}~\bibnamefont {Lei}},
		\bibinfo {author} {\bibfnamefont {F.}~\bibnamefont {Monifi}}, \bibinfo
		{author} {\bibfnamefont {M.}~\bibnamefont {Gianfreda}}, \bibinfo {author}
		{\bibfnamefont {G.~L.}\ \bibnamefont {Long}}, \bibinfo {author}
		{\bibfnamefont {S.}~\bibnamefont {Fan}}, \bibinfo {author} {\bibfnamefont
			{F.}~\bibnamefont {Nori}}, \bibinfo {author} {\bibfnamefont {C.~M.}\
			\bibnamefont {Bender}}, \ and\ \bibinfo {author} {\bibfnamefont
			{L.}~\bibnamefont {Yang}},\ }\bibfield  {title} {\enquote {\bibinfo {title}
			{{Parity–time-symmetric whispering-gallery microcavities}},}\ }\href
	{http://dx.doi.org/10.1038/nphys2927 http://10.0.4.14/nphys2927} {\bibfield
		{journal} {\bibinfo  {journal} {Nat. Phys.}\ }\textbf {\bibinfo {volume}
			{10}},\ \bibinfo {pages} {394} (\bibinfo {year}
		{2014}{\natexlab{a}})}\BibitemShut {NoStop}%
	\bibitem [{\citenamefont {Peng}\ \emph
		{et~al.}(2014{\natexlab{b}})\citenamefont {Peng}, \citenamefont
		{{\"O}zdemir}, \citenamefont {Rotter}, \citenamefont {Yilmaz}, \citenamefont
		{Liertzer}, \citenamefont {Monifi}, \citenamefont {Bender}, \citenamefont
		{Nori},\ and\ \citenamefont {Yang}}]{Peng328}%
	\BibitemOpen
	\bibfield  {author} {\bibinfo {author} {\bibfnamefont {B.}~\bibnamefont
			{Peng}}, \bibinfo {author} {\bibfnamefont {{\c S}.~K.}\ \bibnamefont
			{{\"O}zdemir}}, \bibinfo {author} {\bibfnamefont {S.}~\bibnamefont {Rotter}},
		\bibinfo {author} {\bibfnamefont {H.}~\bibnamefont {Yilmaz}}, \bibinfo
		{author} {\bibfnamefont {M.}~\bibnamefont {Liertzer}}, \bibinfo {author}
		{\bibfnamefont {F.}~\bibnamefont {Monifi}}, \bibinfo {author} {\bibfnamefont
			{C.~M.}\ \bibnamefont {Bender}}, \bibinfo {author} {\bibfnamefont
			{F.}~\bibnamefont {Nori}}, \ and\ \bibinfo {author} {\bibfnamefont
			{L.}~\bibnamefont {Yang}},\ }\bibfield  {title} {\enquote {\bibinfo {title}
			{Loss-induced suppression and revival of lasing},}\ }\href
	{http://science.sciencemag.org/content/346/6207/328} {\bibfield  {journal}
		{\bibinfo  {journal} {Science}\ }\textbf {\bibinfo {volume} {346}},\ \bibinfo
		{pages} {328} (\bibinfo {year} {2014}{\natexlab{b}})}\BibitemShut {NoStop}%
	\bibitem [{\citenamefont {Leefmans}\ \emph {et~al.}(2022)\citenamefont
		{Leefmans}, \citenamefont {Dutt}, \citenamefont {Williams}, \citenamefont
		{Yuan}, \citenamefont {Parto}, \citenamefont {Nori}, \citenamefont {Fan},\
		and\ \citenamefont {Marandi}}]{Leefmans2022}%
	\BibitemOpen
	\bibfield  {author} {\bibinfo {author} {\bibfnamefont {C.}~\bibnamefont
			{Leefmans}}, \bibinfo {author} {\bibfnamefont {A.}~\bibnamefont {Dutt}},
		\bibinfo {author} {\bibfnamefont {J.}~\bibnamefont {Williams}}, \bibinfo
		{author} {\bibfnamefont {L.}~\bibnamefont {Yuan}}, \bibinfo {author}
		{\bibfnamefont {M.}~\bibnamefont {Parto}}, \bibinfo {author} {\bibfnamefont
			{F.}~\bibnamefont {Nori}}, \bibinfo {author} {\bibfnamefont {S.}~\bibnamefont
			{Fan}}, \ and\ \bibinfo {author} {\bibfnamefont {A.}~\bibnamefont
			{Marandi}},\ }\bibfield  {title} {\enquote {\bibinfo {title} {Topological
				dissipation in a time-multiplexed photonic resonator network},}\ }\href
	{http://dx.doi.org/10.1038/s41567-021-01492-w} {\bibfield  {journal}
		{\bibinfo  {journal} {Nat. Phys.}\ }\textbf {\bibinfo {volume} {18}},\
		\bibinfo {pages} {442} (\bibinfo {year} {2022})}\BibitemShut {NoStop}%
	\bibitem [{\citenamefont {Parto}\ \emph {et~al.}(2023)\citenamefont {Parto},
		\citenamefont {Leefmans}, \citenamefont {Williams}, \citenamefont {Nori},\
		and\ \citenamefont {Marandi}}]{Parto2023}%
	\BibitemOpen
	\bibfield  {author} {\bibinfo {author} {\bibfnamefont {M.}~\bibnamefont
			{Parto}}, \bibinfo {author} {\bibfnamefont {C.}~\bibnamefont {Leefmans}},
		\bibinfo {author} {\bibfnamefont {J.}~\bibnamefont {Williams}}, \bibinfo
		{author} {\bibfnamefont {F.}~\bibnamefont {Nori}}, \ and\ \bibinfo {author}
		{\bibfnamefont {A.}~\bibnamefont {Marandi}},\ }\bibfield  {title} {\enquote
		{\bibinfo {title} {Non-{A}belian effects in dissipative photonic topological
				lattices},}\ }\href {\doibase 10.1038/s41467-023-37065-z} {\bibfield
		{journal} {\bibinfo  {journal} {Nat. Commun.}\ }\textbf {\bibinfo {volume}
			{14}},\ \bibinfo {pages} {1440} (\bibinfo {year} {2023})}\BibitemShut
	{NoStop}%
	\bibitem [{\citenamefont {Leefmans}\ \emph {et~al.}(2024)\citenamefont
		{Leefmans}, \citenamefont {Parto}, \citenamefont {Williams}, \citenamefont
		{G.~H. Y.~Li}, \citenamefont {Nori},\ and\ \citenamefont
		{Marandi}}]{Leefmans2024NP}%
	\BibitemOpen
	\bibfield  {author} {\bibinfo {author} {\bibfnamefont {C.~R.}\ \bibnamefont
			{Leefmans}}, \bibinfo {author} {\bibfnamefont {M.}~\bibnamefont {Parto}},
		\bibinfo {author} {\bibfnamefont {J.}~\bibnamefont {Williams}}, \bibinfo
		{author} {\bibfnamefont {A.~Dutt}\ \bibnamefont {G.~H. Y.~Li}}, \bibinfo
		{author} {\bibfnamefont {F.}~\bibnamefont {Nori}}, \ and\ \bibinfo {author}
		{\bibfnamefont {A.}~\bibnamefont {Marandi}},\ }\bibfield  {title} {\enquote
		{\bibinfo {title} {Topological temporally mode-locked laser},}\ }\href
	{\doibase 10.1038/s41567-024-02420-4} {\bibfield  {journal} {\bibinfo
			{journal} {Nat. Phys.}\ }\textbf {\bibinfo {volume} {20}},\ \bibinfo {pages}
		{pages852} (\bibinfo {year} {2024})}\BibitemShut {NoStop}%
	\bibitem [{\citenamefont {Zhang}\ \emph {et~al.}(2018)\citenamefont {Zhang},
		\citenamefont {Peng}, \citenamefont {{\"{O}}zdemir}, \citenamefont {Pichler},
		\citenamefont {Krimer}, \citenamefont {Zhao}, \citenamefont {Nori},
		\citenamefont {Liu}, \citenamefont {Rotter},\ and\ \citenamefont
		{Yang}}]{Zhang2018}%
	\BibitemOpen
	\bibfield  {author} {\bibinfo {author} {\bibfnamefont {J.}~\bibnamefont
			{Zhang}}, \bibinfo {author} {\bibfnamefont {B.}~\bibnamefont {Peng}},
		\bibinfo {author} {\bibfnamefont {Ş.~K.}\ \bibnamefont {{\"{O}}zdemir}},
		\bibinfo {author} {\bibfnamefont {K.}~\bibnamefont {Pichler}}, \bibinfo
		{author} {\bibfnamefont {D.~O.}\ \bibnamefont {Krimer}}, \bibinfo {author}
		{\bibfnamefont {G.}~\bibnamefont {Zhao}}, \bibinfo {author} {\bibfnamefont
			{F.}~\bibnamefont {Nori}}, \bibinfo {author} {\bibfnamefont {Y.~X.}\
			\bibnamefont {Liu}}, \bibinfo {author} {\bibfnamefont {S.}~\bibnamefont
			{Rotter}}, \ and\ \bibinfo {author} {\bibfnamefont {L.}~\bibnamefont
			{Yang}},\ }\bibfield  {title} {\enquote {\bibinfo {title} {A phonon laser
				operating at an exceptional point},}\ }\href
	{https://doi.org/10.1038/s41566-018-0213-5} {\bibfield  {journal} {\bibinfo
			{journal} {Nat. Photon.}\ }\textbf {\bibinfo {volume} {12}},\ \bibinfo
		{pages} {479} (\bibinfo {year} {2018})}\BibitemShut {NoStop}%
	\bibitem [{\citenamefont {Choi}\ \emph {et~al.}(2018)\citenamefont {Choi},
		\citenamefont {Hahn}, \citenamefont {Yoon},\ and\ \citenamefont
		{Song}}]{Choi2018}%
	\BibitemOpen
	\bibfield  {author} {\bibinfo {author} {\bibfnamefont {Y.}~\bibnamefont
			{Choi}}, \bibinfo {author} {\bibfnamefont {C.}~\bibnamefont {Hahn}}, \bibinfo
		{author} {\bibfnamefont {J.~W.}\ \bibnamefont {Yoon}}, \ and\ \bibinfo
		{author} {\bibfnamefont {S.~H.}\ \bibnamefont {Song}},\ }\bibfield  {title}
	{\enquote {\bibinfo {title} {Observation of an anti-{PT}-symmetric
				exceptional point and energy-difference conserving dynamics in electrical
				circuit resonators},}\ }\href {http://dx.doi.org/10.1038/s41467-018-04690-y}
	{\bibfield  {journal} {\bibinfo  {journal} {Nat. Commun.}\ }\textbf {\bibinfo
			{volume} {9}},\ \bibinfo {pages} {2182} (\bibinfo {year} {2018})}\BibitemShut
	{NoStop}%
	\bibitem [{\citenamefont {Helbig}\ \emph {et~al.}(2020)\citenamefont {Helbig},
		\citenamefont {Hofmann}, \citenamefont {Imhof}, \citenamefont {Abdelghany},
		\citenamefont {Kiessling}, \citenamefont {Molenkamp}, \citenamefont {Lee},
		\citenamefont {Szameit}, \citenamefont {Greiter},\ and\ \citenamefont
		{Thomale}}]{Helbig2020}%
	\BibitemOpen
	\bibfield  {author} {\bibinfo {author} {\bibfnamefont {T.}~\bibnamefont
			{Helbig}}, \bibinfo {author} {\bibfnamefont {T.}~\bibnamefont {Hofmann}},
		\bibinfo {author} {\bibfnamefont {S.}~\bibnamefont {Imhof}}, \bibinfo
		{author} {\bibfnamefont {M.}~\bibnamefont {Abdelghany}}, \bibinfo {author}
		{\bibfnamefont {T.}~\bibnamefont {Kiessling}}, \bibinfo {author}
		{\bibfnamefont {L.~W.}\ \bibnamefont {Molenkamp}}, \bibinfo {author}
		{\bibfnamefont {C.~H.}\ \bibnamefont {Lee}}, \bibinfo {author} {\bibfnamefont
			{A.}~\bibnamefont {Szameit}}, \bibinfo {author} {\bibfnamefont
			{M.}~\bibnamefont {Greiter}}, \ and\ \bibinfo {author} {\bibfnamefont
			{R.}~\bibnamefont {Thomale}},\ }\bibfield  {title} {\enquote {\bibinfo
			{title} {Generalized bulk{\textendash}boundary correspondence in
				non-{H}ermitian topolectrical circuits},}\ }\href {\doibase
		10.1038/s41567-020-0922-9} {\bibfield  {journal} {\bibinfo  {journal} {Nat.
				Phys.}\ }\textbf {\bibinfo {volume} {16}},\ \bibinfo {pages} {747} (\bibinfo
		{year} {2020})}\BibitemShut {NoStop}%
	\bibitem [{\citenamefont {Wu}\ \emph {et~al.}(2022)\citenamefont {Wu},
		\citenamefont {Wang}, \citenamefont {Biao}, \citenamefont {Fei},
		\citenamefont {Zhang}, \citenamefont {Yin}, \citenamefont {Hu}, \citenamefont
		{Song}, \citenamefont {Wu}, \citenamefont {Song},\ and\ \citenamefont
		{Yu}}]{wu2022non}%
	\BibitemOpen
	\bibfield  {author} {\bibinfo {author} {\bibfnamefont {J.}~\bibnamefont
			{Wu}}, \bibinfo {author} {\bibfnamefont {Z.}~\bibnamefont {Wang}}, \bibinfo
		{author} {\bibfnamefont {Y.}~\bibnamefont {Biao}}, \bibinfo {author}
		{\bibfnamefont {F.}~\bibnamefont {Fei}}, \bibinfo {author} {\bibfnamefont
			{S.}~\bibnamefont {Zhang}}, \bibinfo {author} {\bibfnamefont
			{Z.}~\bibnamefont {Yin}}, \bibinfo {author} {\bibfnamefont {Y.}~\bibnamefont
			{Hu}}, \bibinfo {author} {\bibfnamefont {Z.}~\bibnamefont {Song}}, \bibinfo
		{author} {\bibfnamefont {T.}~\bibnamefont {Wu}}, \bibinfo {author}
		{\bibfnamefont {F.}~\bibnamefont {Song}}, \ and\ \bibinfo {author}
		{\bibfnamefont {R.}~\bibnamefont {Yu}},\ }\bibfield  {title} {\enquote
		{\bibinfo {title} {Non-{A}belian gauge fields in circuit systems},}\ }\href
	{http://dx.doi.org/10.1038/s41928-022-00833-8} {\bibfield  {journal}
		{\bibinfo  {journal} {Nat. Electron.}\ }\textbf {\bibinfo {volume} {5}},\
		\bibinfo {pages} {635} (\bibinfo {year} {2022})}\BibitemShut {NoStop}%
	\bibitem [{\citenamefont {Zou}\ \emph {et~al.}(2021)\citenamefont {Zou},
		\citenamefont {Chen}, \citenamefont {He}, \citenamefont {Bao}, \citenamefont
		{Lee}, \citenamefont {Sun},\ and\ \citenamefont {Zhang}}]{Zou2021}%
	\BibitemOpen
	\bibfield  {author} {\bibinfo {author} {\bibfnamefont {D.}~\bibnamefont
			{Zou}}, \bibinfo {author} {\bibfnamefont {T.}~\bibnamefont {Chen}}, \bibinfo
		{author} {\bibfnamefont {W.}~\bibnamefont {He}}, \bibinfo {author}
		{\bibfnamefont {J.}~\bibnamefont {Bao}}, \bibinfo {author} {\bibfnamefont
			{C.~H.}\ \bibnamefont {Lee}}, \bibinfo {author} {\bibfnamefont
			{H.}~\bibnamefont {Sun}}, \ and\ \bibinfo {author} {\bibfnamefont
			{X.}~\bibnamefont {Zhang}},\ }\bibfield  {title} {\enquote {\bibinfo {title}
			{Observation of hybrid higher-order skin-topological effect in
				non-{H}ermitian topolectrical circuits},}\ }\href
	{https://doi.org/10.1038/s41467-021-26414-5} {\bibfield  {journal} {\bibinfo
			{journal} {Nat. Commun.}\ }\textbf {\bibinfo {volume} {12}},\ \bibinfo
		{pages} {7201} (\bibinfo {year} {2021})}\BibitemShut {NoStop}%
	\bibitem [{\citenamefont {Hu}\ \emph {et~al.}(2023)\citenamefont {Hu},
		\citenamefont {Zhang}, \citenamefont {Wang}, \citenamefont {Ouyang},
		\citenamefont {Zhu}, \citenamefont {Jia},\ and\ \citenamefont
		{Chan}}]{Hu2023}%
	\BibitemOpen
	\bibfield  {author} {\bibinfo {author} {\bibfnamefont {J.}~\bibnamefont
			{Hu}}, \bibinfo {author} {\bibfnamefont {R.-Y.}\ \bibnamefont {Zhang}},
		\bibinfo {author} {\bibfnamefont {Y.}~\bibnamefont {Wang}}, \bibinfo {author}
		{\bibfnamefont {X.}~\bibnamefont {Ouyang}}, \bibinfo {author} {\bibfnamefont
			{Y.}~\bibnamefont {Zhu}}, \bibinfo {author} {\bibfnamefont {H.}~\bibnamefont
			{Jia}}, \ and\ \bibinfo {author} {\bibfnamefont {C.~T.}\ \bibnamefont
			{Chan}},\ }\bibfield  {title} {\enquote {\bibinfo {title} {Non-{H}ermitian
				swallowtail catastrophe revealing transitions among diverse topological
				singularities},}\ }\href {\doibase 10.1038/s41567-023-02048-w} {\bibfield
		{journal} {\bibinfo  {journal} {Nat. Phys.}\ }\textbf {\bibinfo {volume}
			{19}},\ \bibinfo {pages} {1098} (\bibinfo {year} {2023})}\BibitemShut
	{NoStop}%
	\bibitem [{\citenamefont {Zhang}\ \emph {et~al.}(2025)\citenamefont {Zhang},
		\citenamefont {Wu}, \citenamefont {Yan},\ and\ \citenamefont
		{Chen}}]{PhysRevB.111.014304}%
	\BibitemOpen
	\bibfield  {author} {\bibinfo {author} {\bibfnamefont {X.}~\bibnamefont
			{Zhang}}, \bibinfo {author} {\bibfnamefont {C.}~\bibnamefont {Wu}}, \bibinfo
		{author} {\bibfnamefont {M.}~\bibnamefont {Yan}}, \ and\ \bibinfo {author}
		{\bibfnamefont {G.}~\bibnamefont {Chen}},\ }\bibfield  {title} {\enquote
		{\bibinfo {title} {Observation of non-{H}ermitian pseudo-mobility-edge in a
				coupled electric circuit ladder},}\ }\href {\doibase
		10.1103/PhysRevB.111.014304} {\bibfield  {journal} {\bibinfo  {journal}
			{Phys. Rev. B}\ }\textbf {\bibinfo {volume} {111}},\ \bibinfo {pages}
		{014304} (\bibinfo {year} {2025})}\BibitemShut {NoStop}%
	\bibitem [{\citenamefont {Yamamoto}\ \emph {et~al.}(2019)\citenamefont
		{Yamamoto}, \citenamefont {Nakagawa}, \citenamefont {Adachi}, \citenamefont
		{Takasan}, \citenamefont {Ueda},\ and\ \citenamefont
		{Kawakami}}]{PhysRevLett.123.123601}%
	\BibitemOpen
	\bibfield  {author} {\bibinfo {author} {\bibfnamefont {K.}~\bibnamefont
			{Yamamoto}}, \bibinfo {author} {\bibfnamefont {M.}~\bibnamefont {Nakagawa}},
		\bibinfo {author} {\bibfnamefont {K.}~\bibnamefont {Adachi}}, \bibinfo
		{author} {\bibfnamefont {K.}~\bibnamefont {Takasan}}, \bibinfo {author}
		{\bibfnamefont {M.}~\bibnamefont {Ueda}}, \ and\ \bibinfo {author}
		{\bibfnamefont {N.}~\bibnamefont {Kawakami}},\ }\bibfield  {title} {\enquote
		{\bibinfo {title} {Theory of non-{H}ermitian fermionic superfluidity with a
				complex-valued interaction},}\ }\href {\doibase
		10.1103/PhysRevLett.123.123601} {\bibfield  {journal} {\bibinfo  {journal}
			{Phys. Rev. Lett.}\ }\textbf {\bibinfo {volume} {123}},\ \bibinfo {pages}
		{123601} (\bibinfo {year} {2019})}\BibitemShut {NoStop}%
	\bibitem [{\citenamefont {Nakagawa}\ \emph {et~al.}(2020)\citenamefont
		{Nakagawa}, \citenamefont {Tsuji}, \citenamefont {Kawakami},\ and\
		\citenamefont {Ueda}}]{PhysRevLett.124.147203}%
	\BibitemOpen
	\bibfield  {author} {\bibinfo {author} {\bibfnamefont {M.}~\bibnamefont
			{Nakagawa}}, \bibinfo {author} {\bibfnamefont {N.}~\bibnamefont {Tsuji}},
		\bibinfo {author} {\bibfnamefont {N.}~\bibnamefont {Kawakami}}, \ and\
		\bibinfo {author} {\bibfnamefont {M.}~\bibnamefont {Ueda}},\ }\bibfield
	{title} {\enquote {\bibinfo {title} {Dynamical sign reversal of magnetic
				correlations in dissipative {H}ubbard models},}\ }\href {\doibase
		10.1103/PhysRevLett.124.147203} {\bibfield  {journal} {\bibinfo  {journal}
			{Phys. Rev. Lett.}\ }\textbf {\bibinfo {volume} {124}},\ \bibinfo {pages}
		{147203} (\bibinfo {year} {2020})}\BibitemShut {NoStop}%
	\bibitem [{\citenamefont {Hatano}\ and\ \citenamefont
		{Nelson}(1996)}]{PhysRevLett.77.570}%
	\BibitemOpen
	\bibfield  {author} {\bibinfo {author} {\bibfnamefont {N.}~\bibnamefont
			{Hatano}}\ and\ \bibinfo {author} {\bibfnamefont {D.~R.}\ \bibnamefont
			{Nelson}},\ }\bibfield  {title} {\enquote {\bibinfo {title} {Localization
				transitions in non-{H}ermitian quantum mechanics},}\ }\href {\doibase
		10.1103/PhysRevLett.77.570} {\bibfield  {journal} {\bibinfo  {journal} {Phys.
				Rev. Lett.}\ }\textbf {\bibinfo {volume} {77}},\ \bibinfo {pages} {570}
		(\bibinfo {year} {1996})}\BibitemShut {NoStop}%
	\bibitem [{\citenamefont {Hatano}\ and\ \citenamefont
		{Nelson}(1998)}]{PhysRevB.58.8384}%
	\BibitemOpen
	\bibfield  {author} {\bibinfo {author} {\bibfnamefont {N.}~\bibnamefont
			{Hatano}}\ and\ \bibinfo {author} {\bibfnamefont {D.~R.}\ \bibnamefont
			{Nelson}},\ }\bibfield  {title} {\enquote {\bibinfo {title} {Non-{H}ermitian
				delocalization and eigenfunctions},}\ }\href {\doibase
		10.1103/PhysRevB.58.8384} {\bibfield  {journal} {\bibinfo  {journal} {Phys.
				Rev. B}\ }\textbf {\bibinfo {volume} {58}},\ \bibinfo {pages} {8384}
		(\bibinfo {year} {1998})}\BibitemShut {NoStop}%
	\bibitem [{\citenamefont {Feinberg}\ and\ \citenamefont
		{Zee}(1999)}]{PhysRevE.59.6433}%
	\BibitemOpen
	\bibfield  {author} {\bibinfo {author} {\bibfnamefont {J.}~\bibnamefont
			{Feinberg}}\ and\ \bibinfo {author} {\bibfnamefont {A.}~\bibnamefont {Zee}},\
	}\bibfield  {title} {\enquote {\bibinfo {title} {Non-{H}ermitian localization
				and delocalization},}\ }\href {\doibase 10.1103/PhysRevE.59.6433} {\bibfield
		{journal} {\bibinfo  {journal} {Phys. Rev. E}\ }\textbf {\bibinfo {volume}
			{59}},\ \bibinfo {pages} {6433} (\bibinfo {year} {1999})}\BibitemShut
	{NoStop}%
	\bibitem [{\citenamefont {Gong}\ \emph {et~al.}(2018)\citenamefont {Gong},
		\citenamefont {Ashida}, \citenamefont {Kawabata}, \citenamefont {Takasan},
		\citenamefont {Higashikawa},\ and\ \citenamefont {Ueda}}]{PhysRevX.8.031079}%
	\BibitemOpen
	\bibfield  {author} {\bibinfo {author} {\bibfnamefont {Z.}~\bibnamefont
			{Gong}}, \bibinfo {author} {\bibfnamefont {Y.}~\bibnamefont {Ashida}},
		\bibinfo {author} {\bibfnamefont {K.}~\bibnamefont {Kawabata}}, \bibinfo
		{author} {\bibfnamefont {K.}~\bibnamefont {Takasan}}, \bibinfo {author}
		{\bibfnamefont {S.}~\bibnamefont {Higashikawa}}, \ and\ \bibinfo {author}
		{\bibfnamefont {M.}~\bibnamefont {Ueda}},\ }\bibfield  {title} {\enquote
		{\bibinfo {title} {Topological phases of non-{H}ermitian systems},}\ }\href
	{\doibase 10.1103/PhysRevX.8.031079} {\bibfield  {journal} {\bibinfo
			{journal} {Phys. Rev. X}\ }\textbf {\bibinfo {volume} {8}},\ \bibinfo {pages}
		{031079} (\bibinfo {year} {2018})}\BibitemShut {NoStop}%
	\bibitem [{\citenamefont {Kawabata}\ and\ \citenamefont
		{Ryu}(2021)}]{PhysRevLett.126.166801}%
	\BibitemOpen
	\bibfield  {author} {\bibinfo {author} {\bibfnamefont {K.}~\bibnamefont
			{Kawabata}}\ and\ \bibinfo {author} {\bibfnamefont {S.}~\bibnamefont {Ryu}},\
	}\bibfield  {title} {\enquote {\bibinfo {title} {Nonunitary scaling theory of
				non-{H}ermitian localization},}\ }\href {\doibase
		10.1103/PhysRevLett.126.166801} {\bibfield  {journal} {\bibinfo  {journal}
			{Phys. Rev. Lett.}\ }\textbf {\bibinfo {volume} {126}},\ \bibinfo {pages}
		{166801} (\bibinfo {year} {2021})}\BibitemShut {NoStop}%
	\bibitem [{\citenamefont {Jiang}\ \emph {et~al.}(2019)\citenamefont {Jiang},
		\citenamefont {Lang}, \citenamefont {Yang}, \citenamefont {Zhu},\ and\
		\citenamefont {Chen}}]{PhysRevB.100.054301}%
	\BibitemOpen
	\bibfield  {author} {\bibinfo {author} {\bibfnamefont {H.}~\bibnamefont
			{Jiang}}, \bibinfo {author} {\bibfnamefont {L.-J.}\ \bibnamefont {Lang}},
		\bibinfo {author} {\bibfnamefont {C.}~\bibnamefont {Yang}}, \bibinfo {author}
		{\bibfnamefont {S.-L.}\ \bibnamefont {Zhu}}, \ and\ \bibinfo {author}
		{\bibfnamefont {S.}~\bibnamefont {Chen}},\ }\bibfield  {title} {\enquote
		{\bibinfo {title} {Interplay of non-{H}ermitian skin effects and {A}nderson
				localization in nonreciprocal quasiperiodic lattices},}\ }\href {\doibase
		10.1103/PhysRevB.100.054301} {\bibfield  {journal} {\bibinfo  {journal}
			{Phys. Rev. B}\ }\textbf {\bibinfo {volume} {100}},\ \bibinfo {pages}
		{054301} (\bibinfo {year} {2019})}\BibitemShut {NoStop}%
	\bibitem [{\citenamefont {Wang}\ and\ \citenamefont
		{Wang}(2020)}]{PhysRevB.101.165114}%
	\BibitemOpen
	\bibfield  {author} {\bibinfo {author} {\bibfnamefont {C.}~\bibnamefont
			{Wang}}\ and\ \bibinfo {author} {\bibfnamefont {X.~R.}\ \bibnamefont
			{Wang}},\ }\bibfield  {title} {\enquote {\bibinfo {title} {Level statistics
				of extended states in random non-{H}ermitian {H}amiltonians},}\ }\href
	{\doibase 10.1103/PhysRevB.101.165114} {\bibfield  {journal} {\bibinfo
			{journal} {Phys. Rev. B}\ }\textbf {\bibinfo {volume} {101}},\ \bibinfo
		{pages} {165114} (\bibinfo {year} {2020})}\BibitemShut {NoStop}%
	\bibitem [{\citenamefont {Longhi}(2019)}]{PhysRevLett.122.237601}%
	\BibitemOpen
	\bibfield  {author} {\bibinfo {author} {\bibfnamefont {S.}~\bibnamefont
			{Longhi}},\ }\bibfield  {title} {\enquote {\bibinfo {title} {Topological
				phase transition in non-{H}ermitian quasicrystals},}\ }\href {\doibase
		10.1103/PhysRevLett.122.237601} {\bibfield  {journal} {\bibinfo  {journal}
			{Phys. Rev. Lett.}\ }\textbf {\bibinfo {volume} {122}},\ \bibinfo {pages}
		{237601} (\bibinfo {year} {2019})}\BibitemShut {NoStop}%
	\bibitem [{\citenamefont {Tzortzakakis}\ \emph {et~al.}(2020)\citenamefont
		{Tzortzakakis}, \citenamefont {Makris},\ and\ \citenamefont
		{Economou}}]{PhysRevB.101.014202}%
	\BibitemOpen
	\bibfield  {author} {\bibinfo {author} {\bibfnamefont {A.~F.}\ \bibnamefont
			{Tzortzakakis}}, \bibinfo {author} {\bibfnamefont {K.~G.}\ \bibnamefont
			{Makris}}, \ and\ \bibinfo {author} {\bibfnamefont {E.~N.}\ \bibnamefont
			{Economou}},\ }\bibfield  {title} {\enquote {\bibinfo {title}
			{Non-{H}ermitian disorder in two-dimensional optical lattices},}\ }\href
	{\doibase 10.1103/PhysRevB.101.014202} {\bibfield  {journal} {\bibinfo
			{journal} {Phys. Rev. B}\ }\textbf {\bibinfo {volume} {101}},\ \bibinfo
		{pages} {014202} (\bibinfo {year} {2020})}\BibitemShut {NoStop}%
	\bibitem [{\citenamefont {Claes}\ and\ \citenamefont
		{Hughes}(2021)}]{PhysRevB.103.L140201}%
	\BibitemOpen
	\bibfield  {author} {\bibinfo {author} {\bibfnamefont {J.}~\bibnamefont
			{Claes}}\ and\ \bibinfo {author} {\bibfnamefont {Taylor~L.}\ \bibnamefont
			{Hughes}},\ }\bibfield  {title} {\enquote {\bibinfo {title} {Skin effect and
				winding number in disordered non-{H}ermitian systems},}\ }\href {\doibase
		10.1103/PhysRevB.103.L140201} {\bibfield  {journal} {\bibinfo  {journal}
			{Phys. Rev. B}\ }\textbf {\bibinfo {volume} {103}},\ \bibinfo {pages}
		{L140201} (\bibinfo {year} {2021})}\BibitemShut {NoStop}%
	\bibitem [{\citenamefont {Luo}\ \emph {et~al.}(2021)\citenamefont {Luo},
		\citenamefont {Ohtsuki},\ and\ \citenamefont
		{Shindou}}]{PhysRevLett.126.090402}%
	\BibitemOpen
	\bibfield  {author} {\bibinfo {author} {\bibfnamefont {X.}~\bibnamefont
			{Luo}}, \bibinfo {author} {\bibfnamefont {T.}~\bibnamefont {Ohtsuki}}, \ and\
		\bibinfo {author} {\bibfnamefont {R.}~\bibnamefont {Shindou}},\ }\bibfield
	{title} {\enquote {\bibinfo {title} {Universality classes of the {A}nderson
				transitions driven by non-{H}ermitian disorder},}\ }\href {\doibase
		10.1103/PhysRevLett.126.090402} {\bibfield  {journal} {\bibinfo  {journal}
			{Phys. Rev. Lett.}\ }\textbf {\bibinfo {volume} {126}},\ \bibinfo {pages}
		{090402} (\bibinfo {year} {2021})}\BibitemShut {NoStop}%
	\bibitem [{\citenamefont {Wanjura}\ \emph {et~al.}(2021)\citenamefont
		{Wanjura}, \citenamefont {Brunelli},\ and\ \citenamefont
		{Nunnenkamp}}]{PhysRevLett.127.213601}%
	\BibitemOpen
	\bibfield  {author} {\bibinfo {author} {\bibfnamefont {C.~C.}\ \bibnamefont
			{Wanjura}}, \bibinfo {author} {\bibfnamefont {M.}~\bibnamefont {Brunelli}}, \
		and\ \bibinfo {author} {\bibfnamefont {A.}~\bibnamefont {Nunnenkamp}},\
	}\bibfield  {title} {\enquote {\bibinfo {title} {Correspondence between
				non-{H}ermitian topology and directional amplification in the presence of
				disorder},}\ }\href {\doibase 10.1103/PhysRevLett.127.213601} {\bibfield
		{journal} {\bibinfo  {journal} {Phys. Rev. Lett.}\ }\textbf {\bibinfo
			{volume} {127}},\ \bibinfo {pages} {213601} (\bibinfo {year}
		{2021})}\BibitemShut {NoStop}%
	\bibitem [{\citenamefont {Weidemann}\ \emph {et~al.}(2021)\citenamefont
		{Weidemann}, \citenamefont {Kremer}, \citenamefont {Longhi},\ and\
		\citenamefont {A.Szameit}}]{Weidemann2021}%
	\BibitemOpen
	\bibfield  {author} {\bibinfo {author} {\bibfnamefont {S.}~\bibnamefont
			{Weidemann}}, \bibinfo {author} {\bibfnamefont {M.}~\bibnamefont {Kremer}},
		\bibinfo {author} {\bibfnamefont {S.}~\bibnamefont {Longhi}}, \ and\ \bibinfo
		{author} {\bibnamefont {A.Szameit}},\ }\bibfield  {title} {\enquote {\bibinfo
			{title} {Coexistence of dynamical delocalization and spectral localization
				through stochastic dissipation},}\ }\href {\doibase
		10.1038/s41566-021-00823-w} {\bibfield  {journal} {\bibinfo  {journal} {Nat.
				Photon.}\ }\textbf {\bibinfo {volume} {15}},\ \bibinfo {pages} {576}
		(\bibinfo {year} {2021})}\BibitemShut {NoStop}%
	\bibitem [{\citenamefont {Lin}\ \emph {et~al.}(2022)\citenamefont {Lin},
		\citenamefont {Li}, \citenamefont {Xiao}, \citenamefont {Wang}, \citenamefont
		{Yi},\ and\ \citenamefont {Xue}}]{Lin2022}%
	\BibitemOpen
	\bibfield  {author} {\bibinfo {author} {\bibfnamefont {Q.}~\bibnamefont
			{Lin}}, \bibinfo {author} {\bibfnamefont {T.}~\bibnamefont {Li}}, \bibinfo
		{author} {\bibfnamefont {L.}~\bibnamefont {Xiao}}, \bibinfo {author}
		{\bibfnamefont {K.}~\bibnamefont {Wang}}, \bibinfo {author} {\bibfnamefont
			{W.}~\bibnamefont {Yi}}, \ and\ \bibinfo {author} {\bibfnamefont
			{P.}~\bibnamefont {Xue}},\ }\bibfield  {title} {\enquote {\bibinfo {title}
			{Observation of non-{H}ermitian topological {A}nderson insulator in quantum
				dynamics},}\ }\href {\doibase 10.1038/s41467-022-30938-9} {\bibfield
		{journal} {\bibinfo  {journal} {Nat. Commun.}\ }\textbf {\bibinfo {volume}
			{13}},\ \bibinfo {pages} {3229} (\bibinfo {year} {2022})}\BibitemShut
	{NoStop}%
	\bibitem [{\citenamefont {Liu}\ \emph {et~al.}(2023)\citenamefont {Liu},
		\citenamefont {Lu}, \citenamefont {Zhang},\ and\ \citenamefont
		{Jiang}}]{PhysRevB.107.144204}%
	\BibitemOpen
	\bibfield  {author} {\bibinfo {author} {\bibfnamefont {H.}~\bibnamefont
			{Liu}}, \bibinfo {author} {\bibfnamefont {M.}~\bibnamefont {Lu}}, \bibinfo
		{author} {\bibfnamefont {Z.-Q.}\ \bibnamefont {Zhang}}, \ and\ \bibinfo
		{author} {\bibfnamefont {H.}~\bibnamefont {Jiang}},\ }\bibfield  {title}
	{\enquote {\bibinfo {title} {Modified generalized {B}rillouin zone theory
				with on-site disorder},}\ }\href {\doibase 10.1103/PhysRevB.107.144204}
	{\bibfield  {journal} {\bibinfo  {journal} {Phys. Rev. B}\ }\textbf {\bibinfo
			{volume} {107}},\ \bibinfo {pages} {144204} (\bibinfo {year}
		{2023})}\BibitemShut {NoStop}%
	\bibitem [{\citenamefont {Abrahams}\ \emph {et~al.}(1979)\citenamefont
		{Abrahams}, \citenamefont {Anderson}, \citenamefont {Licciardello},\ and\
		\citenamefont {Ramakrishnan}}]{PhysRevLett.42.673}%
	\BibitemOpen
	\bibfield  {author} {\bibinfo {author} {\bibfnamefont {E.}~\bibnamefont
			{Abrahams}}, \bibinfo {author} {\bibfnamefont {P.~W.}\ \bibnamefont
			{Anderson}}, \bibinfo {author} {\bibfnamefont {D.~C.}\ \bibnamefont
			{Licciardello}}, \ and\ \bibinfo {author} {\bibfnamefont {T.~V.}\
			\bibnamefont {Ramakrishnan}},\ }\bibfield  {title} {\enquote {\bibinfo
			{title} {Scaling theory of localization: {A}bsence of quantum diffusion in
				two dimensions},}\ }\href {\doibase 10.1103/PhysRevLett.42.673} {\bibfield
		{journal} {\bibinfo  {journal} {Phys. Rev. Lett.}\ }\textbf {\bibinfo
			{volume} {42}},\ \bibinfo {pages} {673} (\bibinfo {year} {1979})}\BibitemShut
	{NoStop}%
	\bibitem [{\citenamefont {Brouwer}\ \emph {et~al.}(1998)\citenamefont
		{Brouwer}, \citenamefont {Mudry}, \citenamefont {Simons},\ and\ \citenamefont
		{Altland}}]{PhysRevLett.81.862}%
	\BibitemOpen
	\bibfield  {author} {\bibinfo {author} {\bibfnamefont {P.~W.}\ \bibnamefont
			{Brouwer}}, \bibinfo {author} {\bibfnamefont {C.}~\bibnamefont {Mudry}},
		\bibinfo {author} {\bibfnamefont {B.~D.}\ \bibnamefont {Simons}}, \ and\
		\bibinfo {author} {\bibfnamefont {A.}~\bibnamefont {Altland}},\ }\bibfield
	{title} {\enquote {\bibinfo {title} {Delocalization in coupled
				one-dimensional chains},}\ }\href {\doibase 10.1103/PhysRevLett.81.862}
	{\bibfield  {journal} {\bibinfo  {journal} {Phys. Rev. Lett.}\ }\textbf
		{\bibinfo {volume} {81}},\ \bibinfo {pages} {862} (\bibinfo {year}
		{1998})}\BibitemShut {NoStop}%
	\bibitem [{\citenamefont {Martens}(2006)}]{PhysRevLett.96.076603}%
	\BibitemOpen
	\bibfield  {author} {\bibinfo {author} {\bibfnamefont {H.~C.~F.}\
			\bibnamefont {Martens}},\ }\bibfield  {title} {\enquote {\bibinfo {title}
			{Delocalization in weakly coupled disordered wires: {A}pplication to
				conjugated polymers},}\ }\href {\doibase 10.1103/PhysRevLett.96.076603}
	{\bibfield  {journal} {\bibinfo  {journal} {Phys. Rev. Lett.}\ }\textbf
		{\bibinfo {volume} {96}},\ \bibinfo {pages} {076603} (\bibinfo {year}
		{2006})}\BibitemShut {NoStop}%
	\bibitem [{\citenamefont {Weinmann}\ and\ \citenamefont
		{Evangelou}(2014)}]{PhysRevB.90.155411}%
	\BibitemOpen
	\bibfield  {author} {\bibinfo {author} {\bibfnamefont {D.}~\bibnamefont
			{Weinmann}}\ and\ \bibinfo {author} {\bibfnamefont {S.~N.}\ \bibnamefont
			{Evangelou}},\ }\bibfield  {title} {\enquote {\bibinfo {title}
			{Parity-dependent localization in $n$ strongly coupled chains},}\ }\href
	{\doibase 10.1103/PhysRevB.90.155411} {\bibfield  {journal} {\bibinfo
			{journal} {Phys. Rev. B}\ }\textbf {\bibinfo {volume} {90}},\ \bibinfo
		{pages} {155411} (\bibinfo {year} {2014})}\BibitemShut {NoStop}%
	\bibitem [{\citenamefont {Bordia}\ \emph {et~al.}(2016)\citenamefont {Bordia},
		\citenamefont {L\"uschen}, \citenamefont {Hodgman}, \citenamefont
		{Schreiber}, \citenamefont {Bloch},\ and\ \citenamefont
		{Schneider}}]{PhysRevLett.116.140401}%
	\BibitemOpen
	\bibfield  {author} {\bibinfo {author} {\bibfnamefont {P.}~\bibnamefont
			{Bordia}}, \bibinfo {author} {\bibfnamefont {H.~P.}\ \bibnamefont
			{L\"uschen}}, \bibinfo {author} {\bibfnamefont {S.~S.}\ \bibnamefont
			{Hodgman}}, \bibinfo {author} {\bibfnamefont {M.}~\bibnamefont {Schreiber}},
		\bibinfo {author} {\bibfnamefont {I.}~\bibnamefont {Bloch}}, \ and\ \bibinfo
		{author} {\bibfnamefont {U.}~\bibnamefont {Schneider}},\ }\bibfield  {title}
	{\enquote {\bibinfo {title} {Coupling identical one-dimensional many-body
				localized systems},}\ }\href {\doibase 10.1103/PhysRevLett.116.140401}
	{\bibfield  {journal} {\bibinfo  {journal} {Phys. Rev. Lett.}\ }\textbf
		{\bibinfo {volume} {116}},\ \bibinfo {pages} {140401} (\bibinfo {year}
		{2016})}\BibitemShut {NoStop}%
	\bibitem [{\citenamefont {Iadecola}\ and\ \citenamefont {\ifmmode
			\check{Z}\else \v{Z}\fi{}nidari\ifmmode~\check{c}\else
			\v{c}\fi{}}(2019)}]{PhysRevLett.123.036403}%
	\BibitemOpen
	\bibfield  {author} {\bibinfo {author} {\bibfnamefont {T.}~\bibnamefont
			{Iadecola}}\ and\ \bibinfo {author} {\bibfnamefont {M.}~\bibnamefont
			{\ifmmode \check{Z}\else \v{Z}\fi{}nidari\ifmmode~\check{c}\else
				\v{c}\fi{}}},\ }\bibfield  {title} {\enquote {\bibinfo {title} {Exact
				localized and ballistic eigenstates in disordered chaotic spin ladders and
				the {Fermi-Hubbard} model},}\ }\href {\doibase
		10.1103/PhysRevLett.123.036403} {\bibfield  {journal} {\bibinfo  {journal}
			{Phys. Rev. Lett.}\ }\textbf {\bibinfo {volume} {123}},\ \bibinfo {pages}
		{036403} (\bibinfo {year} {2019})}\BibitemShut {NoStop}%
	\bibitem [{\citenamefont {Lin}\ and\ \citenamefont
		{Gong}(2024)}]{PhysRevA.109.033310}%
	\BibitemOpen
	\bibfield  {author} {\bibinfo {author} {\bibfnamefont {X.}~\bibnamefont
			{Lin}}\ and\ \bibinfo {author} {\bibfnamefont {M.}~\bibnamefont {Gong}},\
	}\bibfield  {title} {\enquote {\bibinfo {title} {Fate of localization in a
				coupled free chain and a disordered chain},}\ }\href {\doibase
		10.1103/PhysRevA.109.033310} {\bibfield  {journal} {\bibinfo  {journal}
			{Phys. Rev. A}\ }\textbf {\bibinfo {volume} {109}},\ \bibinfo {pages}
		{033310} (\bibinfo {year} {2024})}\BibitemShut {NoStop}%
	\bibitem [{\citenamefont {Lin}\ \emph {et~al.}(2023)\citenamefont {Lin},
		\citenamefont {Gong},\ and\ \citenamefont {Guo}}]{arXiv:2307.01638}%
	\BibitemOpen
	\bibfield  {author} {\bibinfo {author} {\bibfnamefont {X.}~\bibnamefont
			{Lin}}, \bibinfo {author} {\bibfnamefont {M.}~\bibnamefont {Gong}}, \ and\
		\bibinfo {author} {\bibfnamefont {G.-C.}\ \bibnamefont {Guo}},\ }\bibfield
	{title} {\enquote {\bibinfo {title} {From single-particle to many-body
				mobility edges and the fate of overlapped spectra in coupled disorder
				models},}\ }\href@noop {} {\bibfield  {journal} {\bibinfo  {journal}
			{arXiv:2307.01638}\ } (\bibinfo {year} {2023})}\BibitemShut {NoStop}%
	\bibitem [{SMB()}]{SMBCS2024}%
	\BibitemOpen
	\href@noop {} {}\bibinfo {note} {See \uppercase{S}upplemental
		\uppercase{M}aterial at [URL] for (I) Physical mechanism, (II)
		Eigenenergy-resolved \uppercase{IPR} and mcom, (III) Finite-size effects,
		(IV) Biorthogonal inverse participation ratio, (V) Quenched dynamics, (VI)
		Robustness of Anderson delocalization against imperfect anti-symmetric
		disorder, (VII) Dependence of phase diagrams on $\gamma$, (VIII) Coexistence
		region of Anderson and skin-mode localization, (IX) Circuit Implementation of
		the model, and (X) Experimental results of HN-Hermitian coupled chains
		subject to the symmetric disorder, which includes
		Ref.~\cite{PhysRevB.105.075128}.}\BibitemShut {Stop}%
	\bibitem [{\citenamefont {Molignini}\ \emph {et~al.}(2023)\citenamefont
		{Molignini}, \citenamefont {Arandes},\ and\ \citenamefont
		{Bergholtz}}]{PhysRevResearch.5.033058}%
	\BibitemOpen
	\bibfield  {author} {\bibinfo {author} {\bibfnamefont {P.}~\bibnamefont
			{Molignini}}, \bibinfo {author} {\bibfnamefont {O.}~\bibnamefont {Arandes}},
		\ and\ \bibinfo {author} {\bibfnamefont {E.~J.}\ \bibnamefont {Bergholtz}},\
	}\bibfield  {title} {\enquote {\bibinfo {title} {Anomalous skin effects in
				disordered systems with a single non-{H}ermitian impurity},}\ }\href
	{\doibase 10.1103/PhysRevResearch.5.033058} {\bibfield  {journal} {\bibinfo
			{journal} {Phys. Rev. Res.}\ }\textbf {\bibinfo {volume} {5}},\ \bibinfo
		{pages} {033058} (\bibinfo {year} {2023})}\BibitemShut {NoStop}%
	\bibitem [{\citenamefont {\textit{ et al}.}()}]{zenodo.16411878}%
	\BibitemOpen
	\bibfield  {author} {\bibinfo {author} {\bibfnamefont {W.~W.~Jin}\
			\bibnamefont {\textit{ et al}.}},\ }\bibfield  {title} {\enquote {\bibinfo
			{title} {10.5281/zenodo.15331425},}\ }\href {\doibase
		10.5281/zenodo.16411878} {\ 10.5281/zenodo.16411878}\BibitemShut {NoStop}%
	\bibitem [{\citenamefont {Xiao}\ and\ \citenamefont
		{Chan}(2022)}]{PhysRevB.105.075128}%
	\BibitemOpen
	\bibfield  {author} {\bibinfo {author} {\bibfnamefont {Y.-X.}\ \bibnamefont
			{Xiao}}\ and\ \bibinfo {author} {\bibfnamefont {C.~T.}\ \bibnamefont
			{Chan}},\ }\bibfield  {title} {\enquote {\bibinfo {title} {Topology in
				non-{H}ermitian {C}hern insulators with skin effect},}\ }\href {\doibase
		10.1103/PhysRevB.105.075128} {\bibfield  {journal} {\bibinfo  {journal}
			{Phys. Rev. B}\ }\textbf {\bibinfo {volume} {105}},\ \bibinfo {pages}
		{075128} (\bibinfo {year} {2022})}\BibitemShut {NoStop}%
\end{thebibliography}

\begin{thebibliography}{5}%
	\makeatletter
	\providecommand \@ifxundefined [1]{%
		\@ifx{#1\undefined}
	}%
	\providecommand \@ifnum [1]{%
		\ifnum #1\expandafter \@firstoftwo
		\else \expandafter \@secondoftwo
		\fi
	}%
	\providecommand \@ifx [1]{%
		\ifx #1\expandafter \@firstoftwo
		\else \expandafter \@secondoftwo
		\fi
	}%
	\providecommand \natexlab [1]{#1}%
	\providecommand \enquote  [1]{``#1''}%
	\providecommand \bibnamefont  [1]{#1}%
	\providecommand \bibfnamefont [1]{#1}%
	\providecommand \citenamefont [1]{#1}%
	\providecommand \href@noop [0]{\@secondoftwo}%
	\providecommand \href [0]{\begingroup \@sanitize@url \@href}%
	\providecommand \@href[1]{\@@startlink{#1}\@@href}%
	\providecommand \@@href[1]{\endgroup#1\@@endlink}%
	\providecommand \@sanitize@url [0]{\catcode `\\12\catcode `\$12\catcode
		`\&12\catcode `\#12\catcode `\^12\catcode `\_12\catcode `\%12\relax}%
	\providecommand \@@startlink[1]{}%
	\providecommand \@@endlink[0]{}%
	\providecommand \url  [0]{\begingroup\@sanitize@url \@url }%
	\providecommand \@url [1]{\endgroup\@href {#1}{\urlprefix }}%
	\providecommand \urlprefix  [0]{URL }%
	\providecommand \Eprint [0]{\href }%
	\providecommand \doibase [0]{http://dx.doi.org/}%
	\providecommand \selectlanguage [0]{\@gobble}%
	\providecommand \bibinfo  [0]{\@secondoftwo}%
	\providecommand \bibfield  [0]{\@secondoftwo}%
	\providecommand \translation [1]{[#1]}%
	\providecommand \BibitemOpen [0]{}%
	\providecommand \bibitemStop [0]{}%
	\providecommand \bibitemNoStop [0]{.\EOS\space}%
	\providecommand \EOS [0]{\spacefactor3000\relax}%
	\providecommand \BibitemShut  [1]{\csname bibitem#1\endcsname}%
	\let\auto@bib@innerbib\@empty
	\bibitem [{\citenamefont {Creutz}(1999)}]{PhysRevLett.83.2636SM}%
	\BibitemOpen
	\bibfield  {author} {\bibinfo {author} {\bibfnamefont {M.}~\bibnamefont
			{Creutz}},\ }\bibfield  {title} {\enquote {\bibinfo {title} {End states,
				ladder compounds, and domain-wall fermions},}\ }\href {\doibase
		10.1103/PhysRevLett.83.2636} {\bibfield  {journal} {\bibinfo  {journal}
			{Phys. Rev. Lett.}\ }\textbf {\bibinfo {volume} {83}},\ \bibinfo {pages}
		{2636} (\bibinfo {year} {1999})}\BibitemShut {NoStop}%
	\bibitem [{\citenamefont {Abrahams}\ \emph {et~al.}(1979)\citenamefont
		{Abrahams}, \citenamefont {Anderson}, \citenamefont {Licciardello},\ and\
		\citenamefont {Ramakrishnan}}]{PhysRevLett.42.673SM}%
	\BibitemOpen
	\bibfield  {author} {\bibinfo {author} {\bibfnamefont {E.}~\bibnamefont
			{Abrahams}}, \bibinfo {author} {\bibfnamefont {P.~W.}\ \bibnamefont
			{Anderson}}, \bibinfo {author} {\bibfnamefont {D.~C.}\ \bibnamefont
			{Licciardello}}, \ and\ \bibinfo {author} {\bibfnamefont {T.~V.}\
			\bibnamefont {Ramakrishnan}},\ }\bibfield  {title} {\enquote {\bibinfo
			{title} {Scaling theory of localization: {A}bsence of quantum diffusion in
				two dimensions},}\ }\href {\doibase 10.1103/PhysRevLett.42.673} {\bibfield
		{journal} {\bibinfo  {journal} {Phys. Rev. Lett.}\ }\textbf {\bibinfo
			{volume} {42}},\ \bibinfo {pages} {673} (\bibinfo {year} {1979})}\BibitemShut
	{NoStop}%
	\bibitem [{\citenamefont {Hatano}\ and\ \citenamefont
		{Nelson}(1996)}]{PhysRevLett.77.570SM}%
	\BibitemOpen
	\bibfield  {author} {\bibinfo {author} {\bibfnamefont {N.}~\bibnamefont
			{Hatano}}\ and\ \bibinfo {author} {\bibfnamefont {D.~R.}\ \bibnamefont
			{Nelson}},\ }\bibfield  {title} {\enquote {\bibinfo {title} {Localization
				transitions in non-{H}ermitian quantum mechanics},}\ }\href {\doibase
		10.1103/PhysRevLett.77.570} {\bibfield  {journal} {\bibinfo  {journal} {Phys.
				Rev. Lett.}\ }\textbf {\bibinfo {volume} {77}},\ \bibinfo {pages} {570}
		(\bibinfo {year} {1996})}\BibitemShut {NoStop}%
	\bibitem [{\citenamefont {Gong}\ \emph {et~al.}(2018)\citenamefont {Gong},
		\citenamefont {Ashida}, \citenamefont {Kawabata}, \citenamefont {Takasan},
		\citenamefont {Higashikawa},\ and\ \citenamefont
		{Ueda}}]{PhysRevX.8.031079SM}%
	\BibitemOpen
	\bibfield  {author} {\bibinfo {author} {\bibfnamefont {Z.}~\bibnamefont
			{Gong}}, \bibinfo {author} {\bibfnamefont {Y.}~\bibnamefont {Ashida}},
		\bibinfo {author} {\bibfnamefont {K.}~\bibnamefont {Kawabata}}, \bibinfo
		{author} {\bibfnamefont {K.}~\bibnamefont {Takasan}}, \bibinfo {author}
		{\bibfnamefont {S.}~\bibnamefont {Higashikawa}}, \ and\ \bibinfo {author}
		{\bibfnamefont {M.}~\bibnamefont {Ueda}},\ }\bibfield  {title} {\enquote
		{\bibinfo {title} {Topological phases of non-{H}ermitian systems},}\ }\href
	{\doibase 10.1103/PhysRevX.8.031079} {\bibfield  {journal} {\bibinfo
			{journal} {Phys. Rev. X}\ }\textbf {\bibinfo {volume} {8}},\ \bibinfo {pages}
		{031079} (\bibinfo {year} {2018})}\BibitemShut {NoStop}%
	\bibitem [{\citenamefont {Xiao}\ and\ \citenamefont
		{Chan}(2022)}]{PhysRevB.105.075128SM}%
	\BibitemOpen
	\bibfield  {author} {\bibinfo {author} {\bibfnamefont {Y.-X.}\ \bibnamefont
			{Xiao}}\ and\ \bibinfo {author} {\bibfnamefont {C.~T.}\ \bibnamefont
			{Chan}},\ }\bibfield  {title} {\enquote {\bibinfo {title} {Topology in
				non-{H}ermitian {C}hern insulators with skin effect},}\ }\href {\doibase
		10.1103/PhysRevB.105.075128} {\bibfield  {journal} {\bibinfo  {journal}
			{Phys. Rev. B}\ }\textbf {\bibinfo {volume} {105}},\ \bibinfo {pages}
		{075128} (\bibinfo {year} {2022})}\BibitemShut {NoStop}%
\end{thebibliography}

%

\end{document}